\documentclass[prd,twocolumn,showpacs,preprintnumbers,amsmath,amssymb]{revtex4}
\usepackage{graphicx}% Include figure files
\usepackage{dcolumn}% Align table columns on decimal point
\usepackage{bm}% bold math
\usepackage{amsmath}

\begin{document}
\preprint{APS/123-QED}

\title{Universal Mass Matrix for Quarks and Leptons and CP violation}
\author{J. Barranco}\email{jbarranc@aei.mpg.de } 
 \affiliation{Max-Planck-Institut f\"ur Gravitationsphysik (Albert-Einstein-Institut),\\ 
 Am M\"uhlenberg 1, D-14476 Golm, Germany}
\author{F. Gonz\'alez Canales}\email{ffelix@fisica.unam.mx},\affiliation{Instituto de 
F\'{\i}sica, Universidad Nacional Aut\'onoma de M\'exico, 04510, M\'exico D.F., M\'exico}, 
\author{A. Mondrag\'on}\email{mondra@fisica.unam.mx}\affiliation{Instituto de F\'{\i}sica, 
Universidad Nacional Aut\'onoma de M\'exico, 04510, M\'exico D.F., M\'exico},

\date{\today}

\begin{abstract}
The measurements of the neutrino and quark mixing angles satisfy the empirical relations
called Quark-Lepton Complementarity. These empirical relations suggest the existence
of a correlation between the mixing matrices of leptons and quarks. In this work, we examine
the possibility that this correlation between the mixing angles of quarks and leptons 
originates in the similar hierarchy of quarks  and charged lepton masses and the seesaw
mechanism type~I, that  gives mass to the Majorana neutrinos. We assume that the similar 
mass hierarchies of charged lepton and quark masses allows us to represent all the mass 
matrices of Dirac fermions in terms of a universal form with four texture zeroes.

Keywords: Flavour symmetries; Quark and lepton masses and mixings; Neutrino masses and 
mixings
\end{abstract}

\pacs{12.15.Ff, 14.60.Pq, 12.15.Hh, 14.60.St}
\maketitle

\section{Introduction}
The neutrino oscillations between different flavour states were measured in a series of 
experiments with atmospheric neutrinos~\cite{Fukuda:1998mi}, solar 
neutrinos~\cite{Cleveland:1998nv}, and neutrinos produced in nuclear 
reactors~\cite{Eguchi:2002dm} and accelerators~\cite{Ahn:2002up}. As a result of the global
combined analysis including all dominant and subdominant oscillation effects, the 
difference of the squared neutrino masses and the mixing angles in the lepton mixing 
matrix, $U_{ _{PMNS} }$, were determined at $1 \sigma$ ($3 \sigma$) confidence 
level~\cite{GonzalezGarcia:2007ib}:
\begin{equation}\label{MCGCdatos}
 \begin{array}{l}
  \Delta m_{ 21 }^{ 2 } = 7.67^{ + 0.22 }_{ - 0.21 }  \left( _{ -0.61 }^{ +0.67 } \right) 
   \times 10^{ -5 }~\textrm{eV}^{2}, \\ \\
  \begin{array}{l} 
   \Delta m_{ 31 }^{ 2 } = \left \{  
    \begin{array}{l} 
     -2.37 \pm 0.15 \left( _{-0.46}^{+0.43} \right) \times 10^{ -3 }~\textrm{eV}^{2}, \\  
     \quad ( m_{ \nu _{ 2 } } > m_{ \nu _{ 1 } } > m_{ \nu _{ 3 } } ). \\ \\
     +2.46 \pm 0.15 \left( _{-0.42}^{+0.47} \right) \times 10^{ -3 }~\textrm{eV}^{2}, \\ 
     \quad ( m_{ \nu _{ 3 } } > m_{ \nu _{ 2 } } > m_{ \nu _{ 1 } } ).
    \end{array} \right. 
  \end{array}
 \end{array}
\end{equation}
{\small \begin{equation}
 \begin{array}{ll}
  \theta_{12}^{ l } = 34.5^{o} \pm 1.4 \left( _{-4.0 }^{+4.8 } \right), &
  \theta_{23}^{ l } = 42.3^{o + 5.1 }_{ \;\; -3.3 }  
   \left( _{ \; \; -7.7 }^{ +11.3 } \right), \\ \\
  \theta_{13}^{ l } = 0.0_{ \; \;-0.0 }^{ o +7.9 } 
   \left( _{-0.0}^{+12.9} \right).
 \end{array}
\end{equation} }
Thus, values of the magnitudes of all nine elements of the lepton mixing matrix,
$U_{ _{ PMNS } }$, at $90\%$ CL, are:
{\small \begin{equation}\label{GG:UPMNS}
 U_{ _{ PMNS } } = \left( \begin{array}{ccc}
  0.80 \rightarrow 0.84 & 0.53 \rightarrow  0.60 & 0.00 \rightarrow  0.17 \\
  0.29 \rightarrow 0.52 & 0.51 \rightarrow  0.69 & 0.61 \rightarrow  0.76 \\
  0.26 \rightarrow 0.50 & 0.46 \rightarrow  0.66 & 0.64 \rightarrow  0.79 
 \end{array} \right).
\end{equation} }
The CHOOZ experiment determined an upper bound for the $\theta_{13}^{ l }$ mixing 
angle~\cite{Apollonio:1999ae}. The latest analyses give the following best
values~\cite{PhysRevLett.103.061804,GonzalezGarcia:2010er}: 
\begin{equation}
 \theta_{13}^{ l } = -0.07_{-0.11 }^{+0.18}
\end{equation} 
and (at 1$\sigma$(3$\sigma$))
\begin{equation}
 \begin{array}{l}
  \theta_{13}^{ l } = 5.6_{-2.7 }^{+3.0}\left(\leq 12.5\right)^{o},\;
  \theta_{13}^{ l } = 5.1_{-3.3 }^{+3.0}\left(\leq 12.0\right)^{o},
 \end{array}
\end{equation}
see also~\cite{Maltoni:2008ka}.
On the other hand, in the last years extensive research has been done in the precise 
determination of the values of the $V_{ _{ CKM } }$ quark mixing matrix elements. The most 
precise fit results for the values of the magnitudes of all nine $CKM$ elements
are~\cite{Amsler:2008zzb}:
{\small \begin{equation}
 \begin{array}{l}
 V_{ _{ CKM } } = \\
 \left( \begin{array}{lll}
   0.97419 \pm .00022 & 0.2257 \pm .0010 & 0.00359 \pm .00016  \\
   0.2256 \pm .0010 & 0.97334 \pm .00023 & 0.0415_{ -.0011 }^{+.0010} \\
   0.00874_{ -.00037 }^{ +.00026 } & 0.0407 \pm .0010  & 
   0.999133_{ -.000043 }^{ +.000044 }
 \end{array} \right)
 \end{array}
\end{equation} }
and the Jarlskog invariant is
\begin{equation}
 J^{q} = \left( 3.05_{ - 0.20 }^{ +0.19 } \right) \times  10^{ -5 }.
\end{equation}
We also have the three angles of the unitarity triangle with the following reported best 
values~\cite{Amsler:2008zzb}:
\begin{equation}
 \alpha = \left( 88_{ -5 }^{ +6 } \right)^{o}, \; 
 \beta =  \left( 21.46 \pm 0.71 \right)^{o}, \; 
 \gamma = \left( 77_{ -32 }^{ +30 } \right)^{o}.
\end{equation}
Each of the elements of the  $V_{ _{ CKM } }$ matrix can be extracted from a large number of
decays and, for the purpose of our analysis, will be considered as independent. Hence, 
current knowledge of the mixing angles for the quark sector can be summarized at 1$\sigma$ 
as~\cite{Amsler:2008zzb}:
\begin{equation}\label{PDGdatosang}
 \begin{array}{l}
  \sin \theta_{12}^{q} = 0.2257 \pm 0.0010, \;  
  \sin \theta_{23}^{q} = 0.0415_{ -0.0011 }^{ +0.0010 }, \\ \\
  \sin \theta_{13}^{q} = 0.00359   \pm 0.00016.
 \end{array}
\end{equation}
The solar mixing angle $\theta^{l}_{12}$ and the correponding mixing angle in the quark sector, 
the Cabibbo angle~$\theta^{q}_{12}$, satisfy an interesting  and intriguing numerical relation 
(at $90~\% $ confidence level)~\cite{Smirnov:2004ju},
\begin{equation}\label{t12}
 \theta^{l}_{12} + \theta^{q}_{12} \approx 45^{o} + 2.5^{o} \pm 1.5^{o},
\end{equation} 
see also~\cite{Smirnov:2009dk}. The equation (\ref{t12}) relates the 1-2 mixing angles in
the quark and lepton sectors, it is commonly known as Quark-Lepton Complementarity relation
(QLC) and, if not accidental, it could imply a quark-lepton symmetry. A second QLC relation 
between the  atmospheric and  2-3 mixing angles, is also 
satisfied~\cite{GonzalezCanales:2009zz},
 \begin{equation}
  \theta_{23}^{ l } + \theta_{ 23 }^{ q }  = \left( 44.67^{+5.1}_{ - 3.3 } \right)^{o} .
 \end{equation}
However, this is not as interesting as (\ref{t12}) because $\theta_{23}^{q}$ is only about
$2^{o}$, and the corresponding QLC relation would be satisfied, within the errors, even 
if the angle $\theta_{23}^{q}$ had been zero, as long as $\theta_{23}^{ l }$ is close to the
maximal value $\pi/4$. A third possible QLC relation is not realized at all, or at least not 
realized in the same way, since it is less than $10^{o}$~\cite{GonzalezCanales:2009zz}.
\begin{equation}\label{t13}
 \theta_{13}^{ l } + \theta_{13}^{q} < 8.1^{o}.
\end{equation}
Equations (\ref{t12})-(\ref{t13}) are known as the extended quark lepton Complementarity, for a 
review see~\cite{Minakata:2005rf}. The extended QLC relations could imply a  quark-lepton 
symmetry~\cite{Minakata:2005rf} or a quark lepton unification~\cite{Frampton:2004ud}. \\
A systematic numerical exploration of all CP conserving textures of the neutrino mass matrix 
compatible the QLC relations and the experimental information on neutrino mixings is given 
in~\cite{Plentinger:2006nb}.

The neutrino oscillations do not provide information about either the absolute mass scale or
if neutrinos are Dirac or Majorana particles~\cite{Camilleri:2008zz}. Thus, one of the most 
fundamental problems of the neutrinos physics is the question of the nature of massive 
neutrinos. A direct way to reveal the nature of massive neutrinos is to investigate 
processes in which the total lepton number is not conserved~\cite{PhysRevD.76.116008}. The
matrix elements for these processes are proportional to the effective Majorana neutrino
masses, which are defined as 
\begin{equation}\label{masa_eff.1}
 \langle m_{ll} \rangle \equiv \sum_{j=1}^{3} m_{ \nu_{j} } U_{lj}^{2} ,\qquad  
 l = e, \mu, \tau,
\end{equation}
where $m_{ \nu_{j} }$ are the neutrino Majorana masses and $U_{lj}$ are the elements of the
lepton mixing matrix.

In this work, we will focus our attention on understanding the nature of the QLC relation,
and finding possible values for the effective Majorana neutrino masses. Thus, we made a 
unified treatment of quarks and leptons, where we assumed that the charged lepton and quark
mass matrices have the same generic form with four texture zeroes from a universal $S_{3}$ 
flavor symmetry and its sequential explicit breaking.

\section{Universal mass matrix with a four zeroes texture}
In particle physics, the imposition of a flavour symmetry has been successful in reducing 
the number of parameters of the Standard Model. Recent flavour symmetry models are reviewed
in~\cite{Ishimori:2010au}; see also the references therein. In particular, a permutational 
$S_{3}$ flavor symmetry and its sequential explicit breaking allows us to take the same 
generic form for the mass matrices of all Dirac fermions, conventionally called the 
generalized Fritzsch ansatz with four texture zeroes~\cite{PhysRevD.61.113002, Fritzsch:1999ee}: 
\begin{equation}\label{T_fritzsch} 
 {\bf M_{ i } = }\left( \begin{array}{ccc} 
  0 & A_{ i } & 0 \\ 
  A^{*}_{ i } & B_{ i } & C_{ i } \\ 
  0 & C_{ i } & D_{ i } \end{array}\right), \quad i=u,d,l,\nu_{ _D }.
\end{equation}
where $B_{ i }$, $C_{ i }$ and $D_{ i }$ are real, while 
$ A_{ i } = \left| A_{ i } \right|e^{ i \phi_{i} }$ with 
$ \phi_{i} = \arg \left \{  A_{ i } \right \} $.

In the most general case, all entries in the Hermitian mass matrix $M_{ i }$ are complex and 
nonvanishing. However, without loss of generality, by means of a common unitary transformation 
of the Dirac fields $\Psi_{ u, \nu_{ _D } }$ and $\Psi_{ d, l }$, it is always possible to 
change to a new flavour basis where the off-diagonal elements
$\left( M_{ i } \right)_{13} = \left( M_{ i } \right)_{31}$ 
vanish~\cite{Fritzsch:1999ee}. The vanishing of the  diagonal elements 
 $ \left( M_{ u, \nu_{ _D } }\right)_{11}$ and $\left( M_{ d, l } \right)_{11}$, 
constrains the physics and allows for the predictions of the Cabibbo angle as funtion of 
the $u$ and $d$-type quark masses in the quark sector and the solar angle as funtion of the 
charged leptons and Majorana neutrinos masses in the leptonic sector in good agreement with 
the experimental values.

Then, in the quark sector, $M_{u}$ and $M_{d}$ totally have four texture zeroes and, in the 
leptonic sector, $M_{ \nu_{ _D } }$ and $M_{e}$, totally have four texture zeroes (here a 
pair of off-diagonal texture zeroes are counted as one zero, due to the Hermiticity of 
$M_{ i }$)~\cite{Fritzsch:1999ee}. Hence, following a common convention we will refer to 
$M_{ i }$ as a generalized Fritzsch ansatz with four texture zeros.

Some reasons to propose the validity of a generalized Fritzsch ansatz with four texture 
zeros as a universal form for the mass matrix of all Dirac fermions in the theory are the 
following:
\begin{enumerate}
 \item The idea of $S_{3}$ flavor symmetry and its explicit breaking has  been succesfully
  realized as a  mass matrix with four texture zeroes in the quark sector to interpret the 
  strong mass hierarchy of up and down type quarks~\cite{Fritzsch:1977za}.
 \item The quark mixing angles and the CP violating phase, appearing in the $V_{ _{CKM} }$
  mixing matrix, were computed as explicit, exact functions of the four quark mass ratios 
  {\small $(m_{u}/m_{t},m_{c}/m_{t},m_{d}/m_{b},m_{s}/m_{b})$,} one symmetry breaking 
  parameter defined as $Z^{1/2} \equiv \frac{ C_{i} }{ B_{ i } }$ and one CP violating phase  
  $\phi_{ _{u-d} }= \phi_{u} - \phi_{d}$. Asuming that $Z_{u} = Z_{d} = Z$, a $\chi^{2}$ 
  fit of the theoretical expresion for $V_{ _{CKM } }^{ th }$ to the experimentally 
  determined $V_{ _{CKM } }^{ exp }$ gave  $Z^{1/2} = \left(\frac{81}{32}\right)^{1/2}$ and 
  $\phi_{ _{u-d} }= 90^{o}$, in good agreement with the experimetal 
  data~\cite{PhysRevD.61.113002}. This agreement with improved as the precision of the 
  experimental data has improved and, now, it is very good~\cite{Amsler:2008zzb}. 
 \item Since the mass spectrum of the charged leptons exhibits a hierarchy similar to the
  quark's one, it would be natural to consider the same $S_{3}$ symmetry and its explicit
  breaking to justify the use of the same generic form with four texture zeroes for the 
  charged lepton mass matrix.
 \item As for the Dirac neutrinos, we have no direct information about the absolute values
  or the relative values of the neutrino masses, but the mass matrix with four texture  
  zeroes can be obtained from an $SO(10)$ neutrino model which describes these the data on 
  neutrino masses and mixings well~\cite{Buchmuller:2001dc}.  Furthermore, from 
  supersymmetry arguments, it would be sensible to assume that the Dirac neutrinos have a 
  mass hierarchy similar to that of the u-quarks and it would be natural to take for the 
  Dirac neutrino mass matrix also a matrix with four texture zeroes.
\end{enumerate}
The Hermitian mass matrix (\ref{T_fritzsch}) may be written in terms of a real symmetric 
matrix $\bar{M}_{ i }$ and a diagonal matrix of phases 
$P_{ i }\equiv \textrm{diag}\left[ 1, e^{ i\phi_{ i } }, e^{ i\phi_{ i } } \right]$ as
follows:  
\begin{equation}\label{Polar:FT}
 M_{ i } = P^{\dagger}_{ i } \bar{M}_{ i } P_{ i }\; .
\end{equation}
The real symetric matrix $\bar{M}_{ i }$ may be brought to diagonal form by means of an
orthogonal transformation,
\begin{equation}\label{Defi:Oreal}
 \bar{M}_{ i } = {\bf O }_{ i } \textrm{diag} \left \{ m_{i1} , m_{i2} , m_{i3}  \right \}
 {\bf O }^{ T }_{ i } ,
\end{equation}
where the $m_{ i }$'s are the eigenvalues of $M_{ i }$ and  ${\bf O }_{ i }$ is a real 
orthogonal matrix. Now computing the invariants of the real symetric matrix 
$\bar{M}_{ i }$, $\textrm{tr}\left \{ \bar{M}_{ i } \right \}$, 
$\textrm{tr}\left \{ \bar{M}_{ i }^{2} \right \}$ and 
$\textrm{det}\left \{ \bar{M}_{ i } \right \}$, we may express the parameters $A_{ i }$,
$B_{ i }$, $C_{ i }$ and $D_{ i }$ occuring in (\ref{T_fritzsch}) in terms of the mass
eigenvalues. In this  way,  we get that the $\bar{M}_{ i }$ matrix $(i=u,d,l,\nu_{ _D })$,  
reparametrized in terms of its eigenvalues and the parameter 
$D_{ i } \equiv 1 - \delta_{ i } $ is
\begin{equation}\label{T_fritzsch:ev}
 \bar{ M  }_{ i } =  \left( \begin{array}{ccc} 
   0 & \sqrt{ \frac{ \widetilde{ m }_{ i1 }   
    \widetilde{ m }_{ i2 } }{ 1 -  \delta_{ i } } } & 0 \\ 
   \sqrt{ \frac{ \widetilde{ m }_{ i1 }   \widetilde{ m }_{ i2 } }{  1 - \delta_{ i } } } & 
   \widetilde{ m }_{ i1 }  - 
   \widetilde{ m }_{ i2 } +  \delta_{ i } & 
   \sqrt{ \frac{\delta_{ i } }{ ( 1 - \delta_{ i } ) } f_{ i1 }  f_{ i2 } }   \\ 
   0 & \sqrt{ \frac{ \delta_{ i } }{ ( 1 - \delta_{ i } ) } f_{ i1 } f_{ i2 } } &  
   1 - \delta_{ i }
 \end{array}\right),
\end{equation}
where  $\widetilde{m}_{i1} = \frac{ m_{i1} }{ m_{i3} }$,  
$\widetilde{m}_{i2} = \frac{ | m_{i2} | }{ m_{i3} }$, 
\begin{equation}\label{fs}
 f_{i1}=1-\widetilde{m}_{i1}-\delta_{i}, \quad  f_{i2} =1+ \widetilde{m}_{i2}  - \delta_{i}. 
\end{equation}
The small parameters $\delta_{i}$ are also functions of the mass ratios and the flavor
symmetry breaking parameter $Z^{1/2}_{i}$~\cite{PhysRevD.61.113002}. The flavor 
symmetry breaking parameter $Z^{1/2}_{i}$, which measures the mixing of singlet and
doublet irreducible representations of $S_{3}$, is defined as the ratio
\begin{equation}\label{def:Z}
 Z^{1/2}_{i} = \frac{ \left( M_{ i } \right)_{23} }{ \left( M_{ i } \right)_{22} }.
\end{equation}
It is related with the parameters $\delta_{i}$ by the following cubic 
equation~\cite{PhysRevD.61.113002}:
\begin{equation}\label{ecu:cubica}
 \begin{array}{l}
  \delta_{i}^{ 3 } - \frac{ 1 }{ Z_{i} + 1 } 
  \left(2 + \widetilde{m}_{i2} - \widetilde{m}_{i1} + 
  \left( 1 + 2 \left( \widetilde{m}_{i2} - \widetilde{m}_{i1} \right) \right)  
    Z_{i} \right) \delta_{i}^{2} + \\ \\
  + \frac{ 1 }{ Z_{i} + 1 } 
   \left(  Z_{i} \left( \widetilde{m}_{i2} - \widetilde{m}_{i1} \right) 
    \left( 2 + \widetilde{m}_{i2} - \widetilde{m}_{i1} \right) 
   + \right. \\ \\ \left. 
   + \left( 1 +  \widetilde{m}_{i2} \right) \left( 1 - \widetilde{m}_{i1} \right)
   \right)\delta_{i}
   + \frac{ Z_{ i } \left( \widetilde{m}_{i2} - \widetilde{m}_{i1} \right)^{2} }{
    Z_{ i } + 1 } = 0.
 \end{array}
\end{equation}
Thus, the small parameter $\delta_{i}$ is obtained as the solution of the cubic equation 
(\ref{ecu:cubica}), which vanishes when $Z_{i}$ vanishes. The last term in the left-hand 
side of (\ref{ecu:cubica}) is equal to the product of the three roots of 
(\ref{ecu:cubica}). Therefore, the root that vanishes when $Z_{i}$ vanishes may be written
as
\begin{equation}
 \delta_{i} = \frac{ Z_{ i } }{ Z_{ i } + 1 } 
 \frac{ \left( \widetilde{m}_{i2} - \widetilde{m}_{i1} \right)^{2} }{
  W _{i}\left( Z \right) }
\end{equation}
where $W _{i}\left( Z \right)$ is the product of the two roots of (\ref{ecu:cubica}) which 
do not vanish when $Z_{i}$ vanishes. The explicit form  of $W _{i}\left( Z \right)$ 
is~\cite{PhysRevD.61.113002}:
\begin{equation}
 \begin{array}{l}
  W _{i}\left( Z \right)  = \left[ p^{3}_{i} + 2 q^{2}_{i} + 2q \sqrt{ p^{3}_{i} +
   q^{2}_{i} } \right]^{ \frac{ 1 }{ 3 } } - | p_{i} |  + \\ 
  + \left[ p^{3}_{i} + 2 q^{2}_{i} - 2q_{i} \sqrt{ p^{3}_{i} + q^{2}_{i} } 
  \right]^{ \frac{ 1 }{ 3 } } + \\
  + \frac{1}{9} \left( Z_{i} \left( 2 \left( \widetilde{m}_{i2} - 
  \widetilde{m}_{i1}\right) + 1 \right) + \left( \widetilde{m}_{i2} - 
  \widetilde{m}_{i1}\right) + 2 \right)^{2} \\
  -\frac{1}{3} \left( \left[ q_{i} + \sqrt{ p^{3}_{i} + q^{2}_{i} } 
  \right]^{ \frac{ 1 }{ 3 } } + 
  \left[ q_{i}- \sqrt{ p^{3}_{i} + q^{2}_{i} } \right]^{ \frac{ 1 }{ 3 } } \right) 
  \times \\ \times 
  \left( Z_{i} \left( 2 \left( \widetilde{m}_{i2} - 
  \widetilde{m}_{i1}\right) + 1 \right) + \left( \widetilde{m}_{i2} - 
  \widetilde{m}_{i1}\right) + 2 \right)
 \end{array}
\end{equation}
with 
\begin{equation}
 \begin{array}{l}
  p_{i} = -\frac{1}{3} \frac{ Z_{ i } }{ Z_{ i } + 1 } \left( Z_{i} 
  \left( 2 \left( \widetilde{m}_{i2} - 
  \widetilde{m}_{i1}\right) + 1 \right) + \widetilde{m}_{i2} - \right. \\ \left.
  \widetilde{m}_{i1} + 2 \right)^{2} + \frac{ 1 }{ Z_{ i } + 1 } 
  \left[ Z_{i} \left( \widetilde{m}_{i2} - \widetilde{m}_{i1} \right) 
  \left( \widetilde{m}_{i2} - \widetilde{m}_{i1} + 
  \right. \right. \\ \left. \left. + 2 \right) 
  \left(1 + \widetilde{m}_{i2} \right) \left( 1 - \widetilde{m}_{i1} \right) \right],
 \end{array}
\end{equation}
\begin{equation}
 \begin{array}{l}
 q_{i} = -\frac{1}{27} \frac{ 1 }{ \left( Z_{ i } + 1 \right)^{ 3 } } \left( Z_{i} \left( 2 
  \left( \widetilde{m}_{i2} - \widetilde{m}_{i1}\right) + 1 \right) + \widetilde{m}_{i2} -
  \right. \\ \left.
  \widetilde{m}_{i1} + 2 \right)^{3} + \frac{1}{6} \frac{ 1 }{ \left(  Z_{ i } + 1 
  \right)^{2} } \left[ Z_{i} \left( \widetilde{m}_{i2} - \widetilde{m}_{i1} \right) 
  \left( \widetilde{m}_{i2} -
  \right. \right. \\ \left. \left. - \widetilde{m}_{i1} + 2 \right) 
  \left(1 + \widetilde{m}_{i2} \right) \left( 1 - \widetilde{m}_{i1} \right) \right]
  \left( Z_{i} \left( 2 \left( \widetilde{m}_{i2} - 
  \right.\right. \right. \\ \left.\left.\left.
  - \widetilde{m}_{i1} \right)+ 1 \right) 
  + \widetilde{m}_{i2} - \widetilde{m}_{i1} + 2 \right).
 \end{array}
\end{equation}
Also, the values allowed for the parameters $\delta_{i}$ are in the following range 
 $ 0 < \delta_{ i } < 1 - \widetilde{m}_{ i1 }$.  \\ 

Now, the entries in the real orthogonal  matrix ${\bf O}$, eq.~(\ref{Defi:Oreal}),  
may also be expressed in terms of the eigenvalues of the mass matrix (\ref{T_fritzsch}) as
{\small \begin{equation}\label{M_ortogonal}
 {\bf O_{i}=}\left(\begin{array}{ccc}
  \left[ \frac{ \widetilde{m}_{i2} f_{i1} }{ {\cal D}_{ i1 } } \right]^{ \frac{1}{2} } & 
 -\left[ \frac{ \widetilde{m}_{i1} f_{i2} }{ {\cal D}_{ i2 } } \right]^{ \frac{1}{2} } & 
  \left[ \frac{ \widetilde{m}_{i1} \widetilde{m}_{i2} \delta_{i} }{ {\cal D}_{i3} }
   \right]^{ \frac{1}{2} } \\
  \left[ \frac{ \widetilde{m}_{i1} ( 1 - \delta_{i} ) f_{i1} }{ {\cal D}_{i1}  } 
   \right]^{ \frac{1}{2} } & 
  \left[ \frac{ \widetilde{m}_{i2} ( 1 - \delta_{i} ) f_{i2} }{ {\cal D}_{i2} } 
   \right]^{ \frac{1}{2} }  &  
  \left[ \frac{ ( 1 - \delta_{i} ) \delta_{i} }{ {\cal D}_{i3} } \right]^{ \frac{1}{2} } \\
 -\left[ \frac{ \widetilde{m}_{i1} f_{i2} \delta_{i} }{ {\cal D}_{i1} } 
   \right]^{ \frac{1}{2} } & 
 -\left[ \frac{ \widetilde{m}_{i2} f_{i1} \delta_{i} }{ {\cal D}_{i2} } 
   \right]^{ \frac{1}{2} } &
  \left[ \frac{ f_{i1} f_{i2} }{ {\cal D}_{i3} } \right]^{ \frac{1}{2} }
 \end{array}\right) ,
\end{equation} }
where,
\begin{equation}\label{Ds}
 \begin{array}{l}
 {\cal D}_{i1} = ( 1 - \delta_{i} )( \widetilde{m}_{i1} + \widetilde{m}_{i2} ) 
  ( 1 - \widetilde{m}_{i1} ),  \\\\
 {\cal D}_{i2} = ( 1 - \delta_{i} )( \widetilde{m}_{i1} + \widetilde{m}_{i2} ) 
  ( 1 + \widetilde{m}_{i2} ), \\\\
 {\cal D}_{i3} = ( 1 - \delta_{i} )( 1 - \widetilde{m}_{i1} )( 1 + \widetilde{m}_{i2} ).
 \end{array}
\end{equation}

\section{SEESAW MECHANISM AND PHASES OF THE LEFT-HANDED NEUTRINO MASS MATRIX}
The left-handed Majorana neutrinos naturally acquire their small masses through an 
effective type I seesaw mechanism  of the form 
\begin{equation}\label{subibajadef}
  M_{ \nu_{L} } = M_{ \nu_{D} } M_{ \nu_R }^{-1} M_{ \nu_{D} }^{T},
\end{equation}
where $M_{ \nu_{D} }$ and $ M_{ \nu_{R} }$ denote the Dirac and right handed Majorana 
neutrino mass matrices, respectively. The symmetry of the mass matrix of the left-handed 
Majorana neutrinos, $M_{\nu_{L}} =M_{\nu_{L}}^{ T }$, and the seesaw mechanism of type I, 
eq. (\ref{subibajadef}), fix the form of the right handed Majorana neutrinos mass matrix, 
$M_{ \nu_{R} }$, which has to be nonsingular and symmetric. 
Further restrictions on $M_{ \nu_{R} }$, follow from requiring that $M_{ \nu_{L} }$ also 
has a texture with four zeroes, as will be shown below. With this purpose in mind, the 
seesaw mechanism, eq.~(\ref{subibajadef}), may be written in a more explicit form as:
\begin{equation}\label{defseesaw1}
 M_{ \nu_{L} } = \frac{ 1 }{ \det \left( M_{ \nu_R } \right) }  
 M_{ \nu_{D} } \textrm{adj} \left( M_{ \nu_R } \right) M_{ \nu_{D} }^{T},
\end{equation}
where $\det ( M_{ \nu_R } ) $ and $\textrm{adj} \left( M_{ \nu_R } \right)$ are the 
determinant and adjugate matrix of $M_{ \nu_R }$, respectively. \\
Now, if we consider the more general form of a complex symmetric matrix of $3 \times 3$  
\begin{equation}\label{Matriz:MR}
 M_{ \nu_R } = \left( \begin{array}{ccc} 
   g_{ \nu_{ _R } } &  a_{ \nu_{ _R } } & e_{ \nu_{ _R } }  \\
   a_{ \nu_{ _R } } &  b_{ \nu_{ _R } } & c_{ \nu_{ _R } } \\
   e_{ \nu_{ _R } } &  c_{ \nu_{ _R } } & d_{ \nu_{ _R } }
 \end{array}\right)
\end{equation}
to represent the right handed Majorana neutrinos mass matrix, we may write 
eq.~(\ref{defseesaw1}) in a more explicit form if we express 
$\det ( M_{ \nu_R } ) $  and $\textrm{adj} \left( M_{ \nu_R } \right)$ in terms of the 
cofactors of the elements of the matrix $M_{ \nu_R }$. Then, 
\begin{equation}
 \det ( M_{ \nu_{ _R } } ) = g_{ \nu_{ _R } } X_{11} - 
 a_{ \nu_{ _R } } X_{ 12 } + e_{ \nu_{ _R } } X_{ 13 }
\end{equation}
and
\begin{equation}\label{subibajagenraltipoI}
 M_{ \nu_{L} } = \frac{ 1 }{ \det \left( M_{ \nu_R } \right) }  
 \left( \begin{array}{ccc} 
   G_{ \nu_{ _L } } &  A_{ \nu_{ _L } } & E_{ \nu_{ _L } }  \\
   A_{ \nu_{ _L } } &  B_{ \nu_{ _L } } & C_{ \nu_{ _L } } \\
   E_{ \nu_{ _L } } &  C_{ \nu_{ _L } } & D_{ \nu_{ _L } }
 \end{array}\right),
\end{equation}
where 
{\small \begin{equation}\label{subibajagenraltipoI-2}
 \begin{array}{l}
  G_{ \nu_{ _L } } = X_{22} A_{ \nu_{ _D } }^{2} ,\\ \\
  A_{ \nu_{ _L } } = -X_{12} | A_{ \nu_{ _D } }|^{2} + X_{ 22 } A_{ \nu_{ _D } } 
  B_{ \nu_{ _D } } - X_{23} A_{ \nu_{ _D } } C_{ \nu_{ _D } } ,\\\\
  B_{ \nu_{ _L } } = X_{11} A_{ \nu_{ _D } }^{*2} + X_{ 22 } B_{ \nu_{ _D } }^{2}
    + X_{ 33 } C_{ \nu_{ _D } }^{ 2 } 
  \\ \quad 
  -2 X_{ 12 } A_{ \nu_{ _D } }^{*} B_{ \nu_{ _D } } 
    + 2 X_{ 13 } A_{ \nu_{ _D } }^{*} C_{ \nu_{ _D } } 
    - 2 X_{23} B_{ \nu_{ _D } } C_{ \nu_{ _D } }, \\ \\
  E_{ \nu_{ _L } } = X_{ 22 } A_{ \nu_{ _D } } C_{ \nu_{ _D } } - 
    X_{ 23 } A_{ \nu_{ _D } } D_{ \nu_{ _D } }, \\ \\
  C_{ \nu_{ _L } } = X_{ 13 } A_{ \nu_{ _D } }^{*} D_{ \nu_{ _D } } 
    - X_{ 12 } A_{ \nu_{ _D } }^{*} C_{ \nu_{ _D } } 
    + X_{ 22 } B_{ \nu_{ _D } } C_{ \nu_{ _D } } \\ \qquad 
    - X_{ 23 } \left( B_{ \nu_{ _D } } D_{ \nu_{ _D } }  + C_{ \nu_{ _D } }^{ 2 } \right) 
    + X_{ 33 } C_{ \nu_{ _D } } D_{ \nu_{ _D } }, \\ \\
  D_{ \nu_{ _L } } = X_{ 22 } C_{ \nu_{ _D } }^{ 2 } 
    - 2 X_{ 23 } C_{ \nu_{ _D } } D_{ \nu_{ _D } } + X_{ 33 } D_{ \nu_{ _D } }^{ 2 }.
 \end{array}
\end{equation} }
In these expresions, the $X_{ nm }$ ($m,n= 1,2,3$) are the cofactors of the correponding 
elements of the $\textrm{adj} \left( M_{ \nu_R } \right)$  
matrix~\footnote{ The cofactors of the elements of  $M_{ \nu_R }$ matrix, are defined as 
$X_{ nm } = ( -1 )^{ n + m } \det \left( H_{nm} \right)$, where $H_{nm}$ is obtained by 
deleting the $n$ row and the $m$ column of $M_{ \nu_R }$ matrix.}.

From eqs.~(\ref{subibajagenraltipoI})~and~(\ref{subibajagenraltipoI-2}), when conditions 
$X_{22} = X_{23}= 0$ are satisfied, the mass matrix of the left-handed Majorana neutrinos will 
have the same universal form with four texture zeroes as the Dirac mass matrices. These 
conditions are equivalent to
\begin{equation}\label{ConFourZeros}
 \begin{array}{l}
  g_{ \nu_{ _R } } d_{ \nu_{ _R } } = e_{ \nu_{ _R } }^{2}, \qquad
  g_{ \nu_{ _R } } c_{ \nu_{ _R } } = a_{ \nu_{ _R } } e_{ \nu_{ _R } },
 \end{array}
\end{equation}
Thus, we obtain the relation 
\begin{equation}
\begin{array}{c}
\frac{ a_{ \nu_{ _R } } }{ c_{ \nu_{ _R } } } = 
   \frac{ e_{ \nu_{ _R } } }{ d_{ \nu_{ _R } } }.
 \end{array}
\end{equation}
For non vanishing $\det ( M_{ \nu_R } )$, these conditions~(\ref{ConFourZeros}) are satisfied,  
if 
\begin{equation}\label{condseesawinv}
 g_{ \nu_{ _R } }=0 \quad \textrm{and} \quad e_{ \nu_{ _R } }=0.
\end{equation}
If we extend the meaning of a mass matrix with four texture zeroes, defined 
in~(\ref{T_fritzsch}), to include the symmetric mass matrix of the right-handed Majorana 
neutrinos, $M_{ \nu_{ _R } }$~\cite{Xing:2003zd}, which is non-Hermitian, we  could say 
that the matrix with four zeroes texture is invariant under the action of the seesaw 
mechanism of type I~\cite{Xing:2003zd,Fritzsch:1999ee, GonzalezCanales:2009zz}. \\
It may also be noticed that, if we set $b_{ \nu_{ _R } } = 0$  or/and 
$c_{ \nu_{ _R } } = 0$, the resulting expression for $M_{ \nu_{ _L } }$ still has four 
texture zeroes. Therefore, $M_{ \nu_{ _L } }$ may also have a  four texture zeroes 
when $M_{ \nu_{ _R } }$ has four, three or two texture zeroes (the two last cases 
are called Fritzsch textures). 

Let us further assume that the phases in the entries of the  $M_{ \nu_{ R} }$  may be 
factorized out as 
\begin{equation}
 M_{ \nu_{ _R } } = R \bar{M}_{ \nu_{ _R } }R,
\end{equation}
where
\begin{equation}
 \bar{M}_{ \nu_{ _R } } = \left( \begin{array}{ccc} 
   0 &  a_{ \nu_{ _R } }  & 0  \\
   a_{ \nu_{ _R } }   & | b_{ \nu_{ _R } } |    &| c_{ \nu_{ _R } } | \\
   0 &  | c_{ \nu_{ _R } }  | & d_{ \nu_{ _R } }   
 \end{array}\right),
\end{equation}
and $R \equiv \textrm{diag}\left[ e^{ - i\phi_{ c }}, e^{ i\phi_{ c } }, 1 \right]$ with 
$\phi_{ c } \equiv \arg \left \{ c_{ \nu_{ _R } }  \right \}$. \\
Then, the type I seesaw mechanism takes the form:
\begin{equation}\label{seesaw9}
  M_{  \nu_{ _L }  } = P_{ _D }^{\dagger}  \bar{M}_{\nu_{ _D } } P_{ _D }
  R^{ \dagger } \bar{M}_{ \nu_{ _R } } ^{-1} R^{ \dagger } 
  P_{ _D } \bar{M}_{\nu_{ _D } } P_{ _D }^{ \dagger },
\end{equation} 
and the mass matrix of the left-handed neutrinos has the following form with four texture
zeroes~\footnote{The seesaw invariance of the four zeroes mass matrix of the Majorana 
neutrino is also derived in~\cite{Xing:2003zd}. However, this authors ignored the phases in 
the elements of  mass matrices in  their  discussion.}: 
\begin{equation}\label{seesaw:F}
  M_{ \nu_{ _L } } =  \left( \begin{array}{ccc} 
   0 &  a_{ \nu_{ _L } } & 0  \\
   a_{ \nu_{ _L } } & b_{ \nu_{ _L } } & c_{ \nu_{ _L } }  \\
   0 & c_{ \nu_{ _L } } & d_{ \nu_{ _L } }   
  \end{array}\right),
\end{equation}
where
{\small \begin{equation}\label{seesaw:F:elem}
 \begin{array}{l}
  a_{ \nu_{ _L } } = \frac{ | a_{ \nu_{ _D } } |^{2} }{  a_{ \nu_{ _R } } } ,\\ \\
  b_{ \nu_{ _L } } = \frac{ c_{ \nu_{ _D } }^{ 2 } }{ d_{ \nu_{ _R } } }  + \frac{ | c_{ 
  \nu_{ _R } }|^{2} -  | b_{ \nu_{ _R } }|  d_{ \nu_{ _R } } }{  d_{ \nu_{ _R } } } 
  \frac{ | a_{ \nu_{ _D } } |^{ 2 } }{  a_{ \nu_{ _R } }^{2} } e^{ i 2\left( \phi_{ c } -
   \phi_{ \nu_{ D } } \right)}   \\
  \qquad  + 2 \frac{ | a_{ \nu_{ _D } }| }{ |a_{ \nu_{ _R } } | } \left( b_{ \nu_{ _D } } 
  e^{- i \phi_{ \nu_{ D } } } 
  - \frac{c_{ \nu_{ _D } } |c_{ \nu_{ _R } }| }{ d_{ \nu_{ _R } }  }  e^{ i \left( \phi_{ c 
  } - \phi_{ \nu_{ D } }   \right) } \right)  ,  \\ \\
  c_{ \nu_{ _L } } = \frac{ c_{ \nu_{ _D } } d_{ \nu_{ _D } } }{ d _{ \nu_{ _R } } }+  
   \\ \quad +
   \frac{ | a_{ \nu_{ _D } } | }{   | a_{ \nu_{ _R } } |}
  \left ( c_{ \nu_{ _D } }e^{- i \phi_{ \nu_{ D } } } - \frac{ |c_{ \nu_{ _R } }| d_{ \nu_{ 
  _D } }  }{ d_{ \nu_{ _R } 
  } } e^{ i \left( \phi_{ c } - \phi_{ \nu_{ D } }  \right) } \right)   , \\ \\
  d_{ \nu_{ _L } } = \frac{ d_{ \nu_{ _D } }^{2} }{ d_{ \nu_{ _R } } }.
 \end{array}
\end{equation} }
The elements $a_{ \nu_{ _L } }$ and $d_{ \nu_{ _L } }$ are real , while 
$b_{ \nu_{ _L } }$ and $c_{ \nu_{ _L } }$ are complex. 
Notice that the phase factors appearing  in eqs. (\ref{seesaw9}) and (\ref{seesaw:F:elem}) 
are fully determined by the seesaw mechanism and our choice of a generalized Fritzsch ansatz 
with four texture  zeroes for the mass matrices of all Dirac fermions and the complex 
symetric, but non-Hermitian, mass matrix of the right handed Majorana neutrinos.

Now, to diagonalize the left-handed Majorana neutrino mass  matrix 
$M_{ \nu_{ _L } }$ by means of a unitary matrix, we need to construct  the hermitian 
matrices $M_{ \nu_{ _L } }M_{ \nu_{ _L } }^{ \dagger }$ and 
$M_{ \nu_{ _L } }^{ \dagger }M_{ \nu_{ _L } }$, which can be diagonalized with unitary
matrices through of the following transformations:
\begin{equation}\label{bilineal:RL}
 \begin{array}{l}
  U_{ _R }^{ \dagger } M_{ \nu_{ _L } }^{ \dagger } M_{ \nu_{ _L } }  U_{ _R } 
   =  \textrm{diag} \left[ \left| m_{ \nu_{1} }^{ s } \right|^{ 2 },  
   \left| m_{ \nu_{2} }^{ s }\right|^{ 2 }, \left| 
   m_{ \nu_{3} }^{ s } \right|^{ 2 }\right], \\ \\
   U_{ _L }^{ \dagger } M_{ \nu_{ _L } } M_{ \nu_{ _L } }^{ \dagger }  U_{ _L }  =  
   \textrm{diag} \left[ \left| 
   m_{ \nu_{1} }^{ s } \right|^{ 2 }, \left| m_{ \nu_{2} }^{ s }\right|^{ 2 },  \left| m_{ 
   \nu_{3} }^{ s }   \right|^{ 2 }\right],
 \end{array}
\end{equation}
where the $ m_{ \nu_{j} }^{ s }$ $(j=1,2,3)$ are the singular values of the 
$M_{ \nu_{ _L } }$ matrix. Thus, with the help of the symmetry of the matrix
(\ref{seesaw:F}) and the transformations (\ref{bilineal:RL}), the left-handed 
Majorana neutrino mass matrix, $M_{ \nu_{ _L } }$, is diagonalized by a unitary matrix 
\begin{equation}
 U_{ \nu }^{ \dagger } M_{ \nu_{ _L } } U_{ \nu }^{ * } = 
 \textrm{diag}\left[ \left| m_{ \nu_{1} }^{ s } \right|,  
 \left| m_{ \nu_{2} }^{ s } \right|, \left| m_{ \nu_{3} }^{ s }\right| \right],
\end{equation}
where $U_{ \nu }\equiv U_{ _L } {\cal K }$ and 
${\cal K } \equiv \textrm{diag}\left[ e^{i\eta_{1}/2}, e^{i\eta_{2}/2}, e^{i\eta_{3}/2} 
\right]$ is the diagonal matrix of the Majorana phases. \\
From the previous analysis, the matrix $M_{ \nu_{ _L } }$ has two non-ignorable phases which 
are 
\begin{equation}\label{fases:ml}
  \phi_{1} \equiv \arg \left\{  b_{ \nu_{ _L } } \right \}  \quad \textrm{and}   \quad   
  \phi_{2} \equiv \arg \left\{ c_{ \nu_{ _L } } \right \}  . 
\end{equation}
However, to discribe the phenomenology of neutrinos masses and mixing, only one phase in 
$ M_{ \nu_{ _L } }$ is required. Therefore, without loss of generality, we may chose 
$\phi_{1}=2\phi_{2}= 2\varphi$ and the following relationship is fulfilled\footnote{The 
general case, when $\phi_{1} \neq 2 \phi_{2}$ is slightly more complicated. This case will 
be trated in detail in a following paper.}:
\begin{equation}
 \tan \phi_{1} = \frac{ 2 \Im m \; c_{ \nu_{ _L } } \Re e \; c_{ \nu_{ _L } } }{ 
   \left(  \Re e \; c_{ \nu_{ _L } }  \right)^{ 2 } - \left( \Im m\;  c_{ \nu_{ _L } } 
   \right)^{ 2 } } .
\end{equation}   
In this case, the analysis simplifies since the phases in $M_{ \nu_{ _L } }$ may be 
factorized out as 
\begin{equation}
 M_{ \nu_{ _L } } = Q \bar{M}_{ \nu_{ _L } } Q,
\end{equation} 
where $Q$ is a diagonal  matrix of phases
{\small $Q \equiv \textrm{diag}\left[ e^{ -i  \varphi }, e^{ i \varphi }, 1 \right] $ }
and $\bar{M}_{ \nu_{ _L } }$ is a real symetric matrix. Then, the matrix $M_{ \nu_{ _L } }$, 
can be diagonalized by a unitary matrix  through the transformation 
\begin{equation}
 U_{ \nu }^{ \dagger } M_{ \nu_{ _L } } U_{ \nu }^{ * } = \textrm{diag}\left[ m_{ \nu_{1} }, 
 m_{ \nu_{2} },  m_{ \nu_{3} } \right];
\end{equation} 
where $m_{ \nu_{j} }$ ($j=1,2,3$) are the eigenvalues of the matrix $M_{ \nu_{ _L } }$, and 
the unitary matrix is 
$U_{ \nu } \equiv Q { \bf O_{\nu} } {\cal K }$  where  $ {\bf O_{\nu} }$ is the orthogonal 
real matrix (\ref{M_ortogonal}), that diagonalizes the real  symetric matrix 
$\bar{M}_{ \nu_{ _L } }$.

It is also important to mention that when the Hermitian matrix with four texture zeroes 
defined in eq. (\ref{T_fritzsch}), is taken as a universal mass matrix for all 
Dirac fermions and right handed Majorana neutrinos~\cite{GonzalezCanales:2009zz}, the phases 
of all entries in the right handed Majorana neutrino mass matrix are fixed at the numerical 
value of $\phi_{ \nu_{ R } } = n \pi$. Thus, the right handed Majorana neutrinos mass matrix 
is real and symmetric and has the form with four texture zeroes shown in (\ref{T_fritzsch}). 
In the more general case in which the Dirac fermions and right handed neutrino mass matrices 
are represented by Hermitian matrices, that can be written in polar form as 
$A=P^{\dagger}\bar{A}P$, where $P$ is a diagonal matrix of phases and $\bar{A}$ is a real 
symmetric matrix, the symmetry of the left-handed Majorana neutrino mass matrix also fixes 
all phases in the mass matrix of the right handed neutrinos at the numerical value 
$\phi_{ \nu_{ R } } = n \pi$. Hence, the only undetermined phases in the mass matrix of the 
left-handed Majorana neutrinos $M_{ \nu_{ _L } }$ are the phases $\phi_{ \nu_{ D } }$, 
coming from the mass matrix of the Dirac neutrinos. 

\section{Mixing Matrices}
The quark and lepton flavor mixing matrices, $U_{ _{PMNS} }$ and $V_{ _{CKM } }$, arise 
from the mismatch between diagonalization of the mass matrices of $u$ and $d$ type 
quarks~\cite{Amsler:2008zzb} and the diagonalization of the mass matrices of
charged leptons and left-handed neutrinos~\cite{Hochmuth:2007wq} respectively,
\begin{equation}\label{M_unitarias}
 U_{ _{PMNS} } = U_{l}^{\dagger}U_{\nu}, \quad  V_{ _{CKM} } = U_{u}U_{d}^{\dagger}.
\end{equation}
Therefore, in order to obtain the unitary matrices appearing in~(\ref{M_unitarias}) and get 
predictions for the flavor mixing angles and CP violating phases, we should specify the
mass matrices. \\
In the quark sector, the unitarity of $V_{ _{CKM} }$ leads to the relations 
$\sum_{i} V_{ij}V_{ik}^{*} = \delta_{jk}$ and $\sum_{j} V_{ij}V_{kj}^{*} = \delta_{ik}$. The
vanishing combinations can be represented as triangles in a complex plane. The area  of all
triangles is equal to half of the Jarlskog invariant, $J_{ q }$~\cite{Jarlskog:1985cw}, 
which is a rephasing invariant measure of CP violation. The term unitarity triangle is 
usually reserved for the tringle obtained from the relation 
$V_{ud}V_{ub}^{*} + V_{cd}V_{cb}^{*} + V_{td}V_{tb}^{*} = 0$. In this case de Jarlskog 
invariant is 
\begin{equation}\label{quarks:JCP}
 J_{ q } = \Im m \left[ V_{us}V_{cs}^{*}V_{ub}^{*}V_{cb} \right], 
\end{equation}
and the inner angles of the unitarity triangle are
\begin{equation}\label{quarks:InerAng}
 \begin{array}{l}
 \alpha \equiv \arg\left( -\frac{ V_{td}V_{tb}^{*} }{ V_{ud}V_{ub}^{*} }\right), \quad 
 \beta \equiv \arg\left( -\frac{ V_{cd}V_{cb}^{*} }{ V_{td}V_{tb}^{*} }\right),  \\ \\
 \gamma \equiv \arg\left( -\frac{ V_{ud}V_{ub}^{*} }{ V_{cd}V_{cb}^{*} }\right).
 \end{array}
\end{equation}
For the lepton sector, when the left-handed neutrinos are Majorana particles, 
the mixing matrix is defined as~\cite{Mohapatra:2006gs} 
$U_{ _{PMNS} } = U_{l}^{\dagger}U_{ _L } K$ where 
$K\equiv \textrm{diag}\left[1, e^{i\beta_{1}}, e^{i\beta_{2}} \right]$ is the diagonal 
matrix of the Majorana CP violating phases. Also in the case of three neutrino mixing there
are three CP violation rephasing invariants~\cite{Hochmuth:2007wq}, associated with the 
three CP violating phases present in the $U_{ _{PMNS} }$ matrix. The rephasing invariant
related to the Dirac phase, analogous to the Jarlskog invariant in the quark sector,
is given by: 
\begin{equation}\label{InvJl}
 J_{ l } \equiv  \Im m \left[ U_{e1}^{ * } U_{ \mu 3 }^{ * } U_{ e3 } U_{ \mu 1 }\right] .
\end{equation} 
The rephasing invariant $J_{ l }$ controls the magnitude of CP violation effects in neutrino 
oscillations and is a directly observable quantity. The other two rephasing invariants 
associated with the two Majorana phases in the $U_{ _{PMNS} }$ matrix, can be chosen as:
\begin{equation}\label{InvS1S2}
 S_{1} \equiv \Im m \left[ U_{e1}U_{ e3 }^{ * }\right], \quad  S_{2} \equiv \Im m \left[ 
 U_{e2}U_{ e3 }^{ * }\right].
\end{equation}
These rephasing invariants are not uniquely defined, but the ones shown in the 
eqs.~(\ref{InvJl})~and~(\ref{InvS1S2}) are relevant for the definition of the effective Majorana 
neutrino mass, $m_{ee}$, in the neutrinoless double beta decay.

\subsection{Mixing Matrices as Functions of the Fermion Masses}
The unitary matrices $U_{u,d}$ occurring in the definition of $V_{ _{CKM} }$, 
eq.~(\ref{M_unitarias}), may be written in polar form as 
$U_{ u,d } =  {\bf O}_{ u,d }^{T} P_{ u,d }$. 
In this expresion, $P_{ u,d }$ is the diagonal matrix of phases appearing in the four texutre 
zeroes mass matrix~(\ref{Polar:FT}). Then, from~(\ref{M_unitarias}), the quark mixing matrix 
takes the form 
\begin{equation}\label{M_unitaria2}
 V_{_{CKM} }^{ ^{th} } =   {\bf O_{u} }^{T} P^{(u-d)}  {\bf O}_{d},
\end{equation}
where $P^{ (u-d) } = \textrm{diag}\left[1, e^{i\phi}, e^{i\phi} \right]$ with $\phi = \phi_{u} 
- \phi_{d}$, and $ {\bf O }_{ u,d }$, are the real orthogonal matrices~(\ref{M_ortogonal}) that 
diagonalize the real symmetric mass matrices $\bar{M}_{i}$. 
A similar analysis shows that $U_{ _{PMNS} }$ may also be written as 
$U_{ _{PMNS} } = U_{l}^{\dagger}U_{\nu}$, with 
$U_{ \nu, l } = P_{ \nu, l } {\bf O_{ \nu, l } }$, this matrix takes the form 
\begin{equation}\label{M_unitaria3}
 U_{ _{PMNS } }^{ ^{th} } = { \bf O}_{l}^{T}P^{ ( \nu - l ) } {\bf O}_{\nu} K,
\end{equation}
where $P^{ ( \nu -l )} = \textrm{diag}\left[1, e^{ i \Phi_{1}  }, e^{ i \Phi_{2} }  \right]$ 
is the diagonal matrix of the Dirac phases, with $\Phi_{1} =  2\varphi - \phi_{ l }$ and 
$\Phi_{2} = \varphi - \phi_{l}$. The real orthogonal matrices ${ \bf O}_{ \nu, l } $ are 
defined in eq.~(\ref{M_ortogonal}).  
Substitution of the expressions (\ref{fs})-(\ref{Ds}) in the unitary matices 
(\ref{M_unitaria2}) and (\ref{M_unitaria3})  allows us to express the mixing 
matrices~$V_{_{CKM}}^{ ^{th} }$~and~$U_{_{PMNS}}^{ ^{th} }$ as explicit functions of the masses 
of quarks and leptons. For the elements of the $V_{_{CKM}}^{ ^{th} }$ mixing matrix, we obtained 
the same theoretical expressions given by Mondrag\'on and 
Rodr\'{\i}guez-Jauregui~\cite{PhysRevD.61.113002}:
\begin{equation}
 V_{_{CKM}}^{ ^{th} } = 
  \left( \begin{array}{ccc}
    V_{ud}^{ ^{th} } &  V_{us}^{ ^{th} } & V_{ub}^{ ^{th} } \\
    V_{cd}^{ ^{th} } &  V_{cs}^{ ^{th} } & V_{cb}^{ ^{th} } \\
    V_{td}^{ ^{th} } &  V_{ts}^{ ^{th} } & V_{tb}^{ ^{th} }
  \end{array} \right),
\end{equation}
where
\begin{widetext}
 \begin{equation}\label{elem:ckm}
  \begin{split}
   \begin{array}{l}
    V_{ ud }^{ ^{th} } = 
     \sqrt{ \frac{ \widetilde{m}_{c} \widetilde{m}_{s} f_{ u1 }  f_{ d1} }{ 
      {\cal D}_{ u 1 } {\cal D}_{ d1 } } } 
      + \sqrt{ \frac{ \widetilde{m}_{u} \widetilde{m}_{d} }{ 
      {\cal D}_{ u 1 } {\cal D}_{ d1 } } } \left( \sqrt{ \left( 1 - \delta_{ u } \right) 
      \left( 1 - \delta_{d} \right) f_{ u1 } f_{ d1 } } + \sqrt{ \delta_{u} \delta_{d} f_{ u2 } 
      f_{ d2 } } \right) e^{ i \phi }, \\
    V_{us}^{ ^{th} } = 
     - \sqrt{ \frac{ \widetilde{m}_{c} \widetilde{m}_{d} f_{ u1 } f_{ d2 } }{ 
      {\cal D}_{ u1 } {\cal D}_{ d2 } } } + \sqrt{ \frac{ \widetilde{m}_{u} \widetilde{m}_{s} }{ 
      {\cal D}_{ u1 } {\cal D}_{ d2 } } } \left( \sqrt{ \left( 1 - \delta_{u} \right) \left( 1 - 
      \delta_{d} \right) f_{ u1 } f_{ d2 }} + \sqrt{ \delta_{u} \delta_{d} f_{ u2 } f_{ d1 } } 
      \right) e^{ i \phi }, \\
    V_{ub}^{ ^{th} } = 
     \sqrt{ \frac{ \widetilde{m}_{c} \widetilde{m}_{d} \widetilde{m}_{s} \delta_{d} f_{ u1 } }{ 
      {\cal D}_{ u1 } {\cal D}_{ d3 } } } + \sqrt{ \frac{ \widetilde{m}_{u} }{ 
      {\cal D}_{ u1 } {\cal D}_{ d3 } } } \left( \sqrt{ \left( 1 - \delta_{u} \right) \left( 1 - 
      \delta_{d} \right) \delta_{d} f_{ u1 } } - \sqrt{ \delta_{u} f_{ u2 } f_{ d1 } f_{ d2 } } 
      \right) e^{ i \phi },\\
    V_{cd}^{ ^{th} } = 
     - \sqrt{ \frac{ \widetilde{m}_{u} \widetilde{m}_{s} f_{ u2 } f_{ d1} }{ 
      {\cal D}_{ u2 } {\cal D}_{ d1 } } } + \sqrt{ \frac{ \widetilde{m}_{c} \widetilde{m}_{d} }{
      {\cal D}_{ u2 } {\cal D}_{ d1 } } } \left( \sqrt{ \left( 1 - \delta_{u} \right) \left( 1 - 
      \delta_{d} \right) f_{ u2 } f_{ d1 } } + \sqrt{ \delta_{u} \delta_{d} f_{ u1 } f_{ d2 } }  
      \right) e^{ i \phi },\\
    V_{cs}^{ ^{th} } = 
     \sqrt{ \frac{ \widetilde{m}_{u} \widetilde{m}_{d} f_{ u2 } f_{ d2} }{ 
      {\cal D}_{ u2 } {\cal D}_{ d2 } } } + \sqrt{ \frac{ \widetilde{m}_{c} \widetilde{m}_{s} }{
      {\cal D}_{ u2 } {\cal D}_{ d2 } } } \left( \sqrt{ \left( 1 - \delta_{u} \right) \left( 1 - 
      \delta_{d} \right) f_{ u2 } f_{ d2 } } + \sqrt{ \delta_{u} \delta_{d} f_{ u1 } f_{ d1 } }  
      \right) e^{ i \phi }, \\
    V_{cb}^{ ^{th} } = 
     - \sqrt{ \frac{ \widetilde{m}_{u} \widetilde{m}_{d} \widetilde{m}_{s} \delta_{d} f_{ u2 } 
      }{ {\cal D}_{ u2 } {\cal D}_{ d3 } } } + \sqrt{ \frac{ \widetilde{m}_{c} }{ 
      {\cal D}_{ u2 } {\cal D}_{ d3 } } }  \left( \sqrt{ \left( 1 - \delta_{u} \right) \left( 1 
      - \delta_{d} \right) \delta_{d} f_{ u2 } } - \sqrt{ \delta_{u} f_{ u1 } f_{ d1 } f_{ d2 } 
      } \right) e^{ i \phi } , \\
   V_{td}^{ ^{th} } = 
    \sqrt{ \frac{ \widetilde{m}_{u} \widetilde{m}_{c} \widetilde{m}_{s} \delta_{u} f_{ d1 } }{ 
     {\cal D}_{ u3 } {\cal D}_{ d1 } } } + \sqrt{ \frac{ \widetilde{m}_{d} }{ 
     {\cal D}_{ u3 } {\cal D}_{ d1 } } } \left( \sqrt{ \delta_{u} \left( 1 - \delta_{u} \right) 
     \left( 1 - \delta_{d} \right) f_{ d1 } } - \sqrt{ \delta_{d} f_{ u1 } f_{ u2 } f_{ d2 } } 
     \right) e^{ i \phi }, \\
  \end{array}
 \end{split}
\end{equation}  
\begin{displaymath}
 \begin{split}
  \begin{array}{l}    
   V_{ts}^{ ^{th} } = 
    - \sqrt{ \frac{ \widetilde{m}_{u} \widetilde{m}_{c} \widetilde{m}_{d} \delta_{u} f_{ d2} }{ 
     {\cal D}_{ u3 } {\cal D}_{ d2 } } } + \sqrt{ \frac{ \widetilde{m}_{s} }{ 
     {\cal D}_{ u3 } {\cal D}_{ d2 } } } \left( \sqrt{ \delta_{u} \left( 1 - \delta_{u} \right) 
     \left( 1 - \delta_{d} \right) f_{ d2 } } - \sqrt{ \delta_{d} f_{ u1 } f_{ u2 } f_{ d1 } } 
     \right) e^{ i \phi },\\
   V_{tb}^{ ^{th} } = 
    \sqrt{ \frac{ \widetilde{m}_{u} \widetilde{m}_{c} \widetilde{m}_{d} \widetilde{m}_{s} 
     \delta_{u} \delta_{d} }{ {\cal D}_{ u3 } {\cal D}_{ d3 } } } + \left( \sqrt{ 
     \frac{ f_{ u1 } f_{ u2 } f_{ d1 } f_{ d2 } }{ {\cal D}_{ u3 } {\cal D}_{ d3 } } } 
     + \sqrt{ \frac{ \delta_{u} \delta_{d} \left( 1 - \delta_{u} \right) \left( 1 - \delta_{d} 
     \right) }{ {\cal D}_{ u3 } D_{ d3 } } } \right) e^{ i \phi }.
  \end{array}
 \end{split}
\end{displaymath}
\end{widetext}
Here, the $m$'s, $f$'s and ${\cal D}$'s are defined in (\ref{fs}) and 
(\ref{Ds}), respectively. And takes the form 
\begin{equation}\label{MsFsDs:quarks}
 \begin{array}{l}
  \widetilde{m}_{u(d)} = \frac{ m_{u(d)} }{ m_{t(b)} },\\
  \widetilde{m}_{c(s)} = \frac{ m_{c(s)} }{ m_{t(b)} },\\
  f_{ u(d)1 } = \left( 1 - \widetilde{m}_{u(d)} - \delta_{u(d)}  \right), \\
  f_{ u(d)2 } = \left( 1 + \widetilde{m}_{c(s)} - \delta_{u(d)}  \right), \\
  {\cal D}_{u(d)1} = ( 1 - \delta_{u(d)} )( \widetilde{m}_{u(d)} + \widetilde{m}_{c(s)} ) 
  ( 1 - \widetilde{m}_{u(d)} ),  \\
 {\cal D}_{u(d)2} = ( 1 - \delta_{u(d)} )( \widetilde{m}_{u(d)} + \widetilde{m}_{c(s)} ) 
  ( 1 + \widetilde{m}_{u(d)} ), \\
 {\cal D}_{u(d)3} = ( 1 - \delta_{u(d)} )( 1 - \widetilde{m}_{u(d)} )( 1 + 
 \widetilde{m}_{c(s)} ).
 \end{array}
\end{equation} 
Now, with the help of the equations (\ref{M_ortogonal}) and (\ref{M_unitaria3}), we obtain 
the theoretical expresion of the elements of the lepton mixing matrix, 
$U_{_{PMNS}}^{ ^{th} }$. This expresions have the following form:
\begin{equation}
 U_{_{ PMNS } }^{ ^{th} } = 
 \left(  \begin{array}{ccc}
  U_{ e 1 }^{ ^{th} } & U_{ e 2 }^{ ^{th} } e^{ i \beta_{ 1 } } &  U_{ e 3 }^{ ^{th} } 
   e^{ i \beta_{ 2 } } \\
  U_{ \mu 1 }^{ ^{th} } & U_{ \mu 2 }^{ ^{th} } e^{ i \beta_{ 1 } } &  U_{ \mu 3 }^{ ^{th} } 
   e^{ i \beta_{ 2 } } \\
  U_{ \tau 1 }^{ ^{th} } & U_{ \tau 2 }^{ ^{th} } e^{ i \beta_{ 1 } } & U_{ \tau 3 }^{ ^{th} } 
   e^{ i \beta_{ 2 } } 
 \end{array} \right)
\end{equation}
where
\begin{widetext}
\begin{equation}\label{elem:pmns}
\begin{split}
 \begin{array}{l}
  U_{e1}^{ ^{th} } = 
   \sqrt{ \frac{\widetilde{m}_{\mu} \widetilde{m}_{\nu_{2}} f_{ l1 } f_{ \nu1 } }{ 
    {\cal D}_{ l1 } {\cal D}_{ \nu1 } } } + \sqrt{ \frac{ \widetilde{m}_{e} 
    \widetilde{m}_{\nu_{1}} }{ {\cal D}_{ l1 } {\cal D}_{ \nu 1 } } } 
    \left( \sqrt{ ( 1 - \delta_{l} )( 1 - \delta_{\nu} ) f_{ l1 } f_{ \nu1 } } 
    e^{ i \Phi_{1} } + \sqrt{ \delta_{l} \delta_{\nu} f_{ l2 } f_{ \nu2 } } e^{ i \Phi_{2} }  
    \right), \\
  U_{e2}^{ ^{th} } = 
   - \sqrt{ \frac{\widetilde{m}_{\mu} \widetilde{m}_{\nu_{1}} f_{ l1 } f_{ \nu2 } }{ 
    {\cal D}_{ l1 } {\cal D}_{ \nu2 } } } + \sqrt{ \frac{ \widetilde{m}_{e} 
    \widetilde{m}_{\nu_{2}} }{ {\cal D}_{ l1 } {\cal D}_{ \nu2 } } } \left( \sqrt{ ( 1 - 
    \delta_{l} )( 1 - \delta_{\nu} ) f_{ l1 }  f_{ \nu2 } } e^{ i \Phi_{ 1 } } +
    \sqrt{ \delta_{l} \delta_{\nu} f_{ l2 }  f_{ \nu1 } } e^{ i \Phi_{2} } \right) , \\
  U_{e3}^{ ^{th} }  = 
   \sqrt{ \frac{ \widetilde{m}_{\mu} \widetilde{m}_{\nu_{1}} \widetilde{m}_{\nu_{2}}   
    \delta_{\nu} f_{ l1 } }{ {\cal D}_{ l1 } {\cal D}_{ \nu3} } } + \sqrt{ \frac{ 
    \widetilde{m}_{e} }{ {\cal D}_{ l1 } {\cal D}_{ \nu3 } } } \left( \sqrt{ 
    \delta_{\nu} ( 1 - \delta_{l} ) ( 1 - \delta_{\nu} ) f_{ l1 } } e^{ i \Phi_{1} }
    - \sqrt{ \delta_{e}  f_{ l2 }  f_{ \nu1 }  f_{ \nu2 } }  e^{ i  \Phi_{2}  }\right) , \\
  U_{ \mu1 }^{ ^{th} } = 
   -\sqrt{ \frac{ \widetilde{m}_{e} \widetilde{m}_{\nu_{2}} f_{ l2 } f_{ \nu1 } }{ 
    {\cal D}_{ l2 } {\cal D}_{ \nu1 } } } + \sqrt{ \frac{ \widetilde{m}_{\mu} 
    \widetilde{m}_{\nu_{1}} }{ {\cal D}_{ l2 } {\cal D}_{ \nu1 } } } \left( \sqrt{ ( 1 - 
    \delta_{l} )( 1 - \delta_{\nu} ) f_{ l2 } f_{ \nu1 } } e^{ i \Phi_{1} } + \sqrt{ 
    \delta_{l} \delta_{\nu} f_{ l1 }  f_{ \nu2 } } e^{ i \Phi_{2} } \right) ,\\
  U_{ \mu2 }^{ ^{th} } = 
   \sqrt{ \frac{ \widetilde{m}_{e} \widetilde{m}_{\nu_{1} } f_{ l2 } f_{ \nu2 } }{ 
    {\cal D}_{ l2 } {\cal D}_{\nu 2} } } + \sqrt{ \frac{ \widetilde{m}_{\mu} 
    \widetilde{m}_{\nu_{2}} }{ {\cal D}_{ l2 } {\cal D}_{ \nu2 } } } \left( \sqrt{ ( 1 -
    \delta_{l} ) ( 1 - \delta_{\nu} ) f_{ l2 } f_{ \nu2 } } e^{ i \Phi_{1} } + \sqrt{
    \delta_{l} \delta_{\nu} f_{ l1 } f_{ \nu1 } } e^{ i \Phi_{2}  } \right ) , \\
  U_{ \mu3 }^{ ^{th} } = 
   -\sqrt{ \frac{\widetilde{m}_{e} \widetilde{m}_{\nu_{1}} \widetilde{m}_{\nu_{2}} 
    \delta_{\nu} f_{ l2 } }{ {\cal D}_{ l2 } {\cal D}_{ \nu3 } } } + \sqrt{ 
    \frac{ \widetilde{m}_{\mu} }{ {\cal D}_{ l2 } {\cal D}_{ \nu3 } } } \left( \sqrt{ 
    \delta_{\nu} ( 1 - \delta_{l} ) ( 1 - \delta_{\nu} ) f_{ l2 } } e^{ i \Phi_{1} }
    - \sqrt{ \delta_{l} f_{ l1 } f_{ \nu1 } f_{ \nu2 } } e^{ i \Phi_{2}  } \right ), \\
  U_{ \tau1 }^{ ^{th} } = 
   \sqrt{ \frac{ \widetilde{m}_{e} \widetilde{m}_{\mu} \widetilde{m}_{\nu_{2}} \delta_{l} 
    f_{ \nu1 } }{ {\cal D}_{ l3 } {\cal D}_{ \nu1 } } } + \sqrt{ \frac{ \widetilde{m}_{\nu_{1}} 
    }{ {\cal D}_{ l3 } {\cal D}_{ \nu1 } } } \left( \sqrt{ \delta_{l} ( 1 - \delta_{l} )( 1 - 
    \delta_{\nu} ) f_{ \nu1 } } e^{ i \Phi_{1} } - \sqrt{ \delta_{ \nu } f_{ l1 } f_{ l2 }  
    f_{ \nu2 } } e^{ i \Phi_{2} } \right), \\
  U_{ \tau2 }^{ ^{th} } =  
   - \sqrt{ \frac{ \widetilde{m}_{e} \widetilde{m}_{\mu} \widetilde{m}_{\nu_{1}} \delta_{l} 
    f_{ \nu2 } }{ {\cal D}_{ l3 } {\cal D}_{\nu 2} } } + \sqrt{ \frac{ \widetilde{m}_{\nu_{2}} 
    }{ {\cal D}_{ l3 } {\cal D}_{ \nu2 } } } \left( \sqrt{ \delta_{l} ( 1 - \delta_{l} )( 1 - 
    \delta_{\nu} ) f_{ \nu2 } } e^{ i \Phi_{1} } - \sqrt{ \delta_{\nu} f_{ l1 } f_{ l2 } 
    f_{ \nu1 } } e^{ i \Phi_{2}  } \right ), \\ 
  U_{ \tau3 }^{ ^{th} } = 
   \sqrt{ \frac{ \widetilde{m}_{e} \widetilde{m}_{\mu} \widetilde{m}_{\nu_{1}} 
    \widetilde{m}_{\nu_{2}} \delta_{l} \delta_{\nu} }{ {\cal D}_{ l3 } {\cal D}_{ \nu3 } } } + 
    \sqrt{ \frac{ \delta_{l} \delta_{\nu}( 1 - \delta_{l} ) ( 1 - \delta_{\nu} ) }{ 
    {\cal D}_{ l3 } {\cal D}_{ \nu3 } } } e^{ i \Phi_{1} } + \sqrt{ \frac{  f_{l1 } f_{ l2 }  
    f_{ \nu1 }  f_{ \nu2 } }{ {\cal D}_{ l3 } {\cal D}_{ \nu3 } } } e^{ i \Phi_{2}  } ,
 \end{array}
 \end{split}
\end{equation}
\end{widetext}
in these expresions the $\widetilde{m}$'s, $f$'s and ${\cal D}$'s are defined in (\ref{fs}) and 
(\ref{Ds}), respectively. And takes the form
{\small \begin{equation}\label{MsFsDs:leptones}
 \begin{array}{l}
  \widetilde{m}_{\nu_{1}(e)} = \frac{ m_{\nu_{1}(e)} }{ m_{\nu_{3}(\tau)} },\\
  \widetilde{m}_{\nu_{2}(\mu)} = \frac{ m_{\nu_{2}(\mu)} }{ m_{\nu_{3}(\tau)} },\\
  f_{ \nu(l)1 } = \left( 1 - \widetilde{m}_{\nu_{1}(e)} - \delta_{\nu(l)}  \right), \\
  f_{ \nu(l)2 } = \left( 1 + \widetilde{m}_{\nu_{2}(\mu)} - \delta_{\nu(l)}  \right), \\
  {\cal D}_{\nu(l)1} = ( 1 - \delta_{\nu(l)} )( \widetilde{m}_{\nu_{1}(e)} + 
  \widetilde{m}_{\nu_{2}(\mu)} ) ( 1 - \widetilde{m}_{\nu_{1}(e)} ),  \\
 {\cal D}_{ \nu(l)2} = ( 1 - \delta_{\nu(l)} )( \widetilde{m}_{\nu_{1}(e)} + 
  \widetilde{m}_{\nu_{2}(\mu)} ) ( 1 + \widetilde{m}_{\nu_{2}(\mu)} ), \\
 {\cal D}_{\nu(l)3} = ( 1 - \delta_{\nu(l)} )( 1 - \widetilde{m}_{\nu_{1}(e)} )( 1 + 
 \widetilde{m}_{\nu_{2}(\mu)} ).
 \end{array}
\end{equation} }

\subsection{The  $\chi^{2}$ fit for the Quark Mixing Matrix}
We made a $\chi^{2}$ fit of the exact theoretical expressions for the modulii of the entries 
of the quark mixing matrix $| ( V_{ _{ CKM } }^{ ^{th} } )_{ij} |$ and the inner angles of the 
unitarity triangle $\alpha^{ ^{th} }$, $\beta^{ ^{th} }$ and $\gamma^{ ^{th} }$ to the 
experimental values given by Amsler~\cite{Amsler:2008zzb}. In this fit, we computed the modulii 
of the entries of the quark mixing matrix and the inner angles of the unitarity triangle from 
the theoretical expresion (\ref{elem:ckm}) with the following numerical values of the quark mass 
ratios~\cite{Amsler:2008zzb}:
\begin{equation}\label{quark-rat}
 \begin{array}{l}
  \widetilde{m}_{u} = 2.5469 \times 10^{ -5}, \quad  
  \widetilde{m}_{c} = 3.9918 \times 10^{ -3}, \\
  \widetilde{m}_{d} = 1.5261 \times 10^{ -3}, \quad  
  \widetilde{m}_{s} = 3.2319 \times 10^{ -2}.
 \end{array}
\end{equation} 
The numerical values of the mass ratios were left fixed at the values given in 
eq.~(\ref{quark-rat}) and the parameters $\delta_{u}$ and $\delta_{d}$ were left as 
free parameters to be varied. Hence, in the $\chi^{2}$ fit we have six degrees of freedom 
($d.o.f.$), namely, the nine observable modulii of the entries in the $V_{ _{CKM} }$ 
matrix  less the three free parameters to be varied. Once the best values of the parameters 
$\delta_{u}$, $\delta_{d}$ and $\phi$ were determined , we computed the three inner angles 
of the unitary triangle from eq.~(\ref{quarks:InerAng}) and the Jarlskog invariant from 
eq.~(\ref{quarks:JCP}).

The resulting best values of the parameters $\delta_{u}$ and $\delta_{d}$ are
\begin{equation}\label{X2:Qds}
 \delta_{u} = 3.829 \times 10^{-3}, \quad \delta_{d} = 4.08 \times 10^{ -4 }  
\end{equation}   
and the Dirac CP violating phase is $\phi = 90^{o}$. The best values for the moduli of the entries of the $CKM$ mixing matrix are given in the 
following expresion
\begin{equation}
 \left| V_{ _{CKM} }^{ ^{th} } \right| =  \left(\begin{array}{ccc}
  0.97421 & 0.22560 & 0.003369 \\
  0.22545 & 0.97335 & 0.041736 \\
  0.008754 & 0.04094 & 0.99912  
 \end{array} \right)
\end{equation}
and inner angles of the unitary triangle
\begin{equation}
 \alpha^{ ^{th} } = 91.24^{o}, \quad \beta^{ ^{th} } = 20.41^{o}, \quad  
 \gamma^{ ^{th} } = 68.33^{o}.
\end{equation}
The Jarlskog invariant takes the value 
\begin{equation}
 J_{q}^{ ^{th} } = 2.9 \times 10^{-5}.
\end{equation}
All these results are in good agreement with the experimental values.  The minimun value of 
$\chi^{2}$ obtained in this fit is 4.6 and the resulting value of $\chi^{2}$ for degree of 
freedom is {\small $\frac{\chi^{2}_{min} }{ d.o.f. }=0.77$}.

\subsection{ The $\chi^{2}$ fit for the Lepton Mixing Matrix}
In the case of the lepton mixing matrix, we made a $\chi^{2}$ fit of the theoretical 
expressions for the modulii of the entries of the lepton mixing matrix 
$| ( U_{ _{ PMNS } }^{ ^{th} } )_{ij} |$ given in eq.~(\ref{elem:pmns}) to the  values 
extracted from experiment as given by Gonzalez-Garcia~\cite{GonzalezGarcia:2007ib} and 
quoted in eq.~(\ref{GG:UPMNS}).The computation was made using the following values for the 
charged lepton masses~\cite{Amsler:2008zzb}:
\begin{equation}\label{massChL}
\begin{array}{l}
 m_{e} = 0.5109~\textrm{MeV}, \;\; m_{\mu}= 105.685~\textrm{MeV}, \;\;   \\ 
 m_{\tau}=1776.99~\textrm{MeV}.
\end{array} 
\end{equation}
We took for the masses of the left-handed Majorana neutrinos a normal hierarchy. This allows 
us to write the left-handed Majorana neutrinos mass ratios in terms of the neutrino 
squared mass differences and  the neutrino mass $m_{ \nu_{3} }$ in the following form:
\begin{equation}
 \begin{array}{l}
 \widetilde{m}_{ \nu_{1} } = \sqrt{ 1 - \frac{ \left( \Delta m_{ 32 }^{ 2 } + 
 \Delta m_{ 21 }^{ 2 } \right) }{
  m_{ \nu_{3} }^{ 2 } } }, \;
 \widetilde{m}_{ \nu_{2} } = 
 \sqrt{ 1 - \frac{ \Delta m_{ 32 }^{ 2 } }{ m_{ \nu_{3} }^{ 2 } } }.
 \end{array}
\end{equation}
The neutrino squared mass differences were obtained from the experimental data on neutrino 
oscillations given in Gonzalez-Garcia~\cite{GonzalezGarcia:2007ib} and we left the mass 
$m_{ \nu_{3} }$ as a free parameter of the $\chi^{2}$ fit. Also, the  parameters  
$\delta_{e}$, $\delta_{\nu}$,  $\Phi_{1}$ and  $\Phi_{2} $ were left as frees parameters to 
be varied. Hence, in this $\chi^{2}$ fit we have four degrees of freedom. \\
 From the best values obtained for $m_{ \nu_{3} }$ and the experimental values of the 
 $\Delta m_{ 32 }^{ 2 }$ and $\Delta m_{ 21 }^{ 2 }$, we obtained the following best 
 values for the neutrino masses
\begin{equation}\label{X2:Mnus}
 \begin{array}{l}
 m_{\nu_{1}} = 2.7  \times 10^{-3}\textrm{eV}, \quad  m_{\nu_{2}} =  9.1  \times 
 10^{-3}\textrm{eV}, \\  m_{\nu_{3}} = 4.7 \times 10^{-2}\textrm{eV}. 
 \end{array}
\end{equation} 
The resulting best values of the parameters $\delta_{e}$ and $\delta_{\nu}$ are 
\begin{equation}\label{X2:ds}
 \delta_{l} = 0.06, \qquad \delta_{ \nu } = 0.522 ,
\end{equation}
and the best values of the Dirac CP violating phases are 
$\Phi_{1} = \pi \quad \textrm{and} \quad \Phi_{2} = 3\pi/2$.
The best values for the modulii of the entries of the $PMNS$ mixing matrix are given in the 
following expresion
\begin{equation}
 \left| U_{ _{PMNS} }^{ ^{th} } \right| =
 \left(\begin{array}{ccc}
  0.820421 & 0.568408 & 0.061817 \\
  0.385027 & 0.613436 & 0.689529 \\
  0.422689 & 0.548277 & 0.721615
 \end{array} \right).
\end{equation} 
The value of the rephasing invariant related to the Dirac phase is 
\begin{equation}
 J_{ l }^{ ^{th} } = 8.8 \times 10^{ -3}. 
\end{equation}
In the absence of experimental information about the Majorana phases 
$\beta_{1 }$ and $\beta_{ 2 }$, the two rephasing invariants $S_{1}$ and $S_{2}$, 
eq.~(\ref{InvS1S2}), associated with the two Majorana phases in the $U_{ _{PMNS} }$ matrix,  
could not be determined from experimental values. Therefore, in order to make a numerical 
estimate of Majorana phases, we maximized the rephasing invariants $S_{1}$ and $S_{2}$, thus 
obtaining a numerical value for the  Majorana phases $\beta_{1 }$ and $\beta_{ 2 }$. Then, 
the maximum values of the rephasing invariants,  
eq(\ref{InvS1S2}), are:
\begin{equation}\label{valor:S1S2}
 S_{1}^{ max } = -4.9 \times 10^{ -2 }, \quad S_{2}^{ max } = 3.4 \times 10^{ -2 },
\end{equation}
with  $\beta_{1 }= -1.4^{ o }$ and $\beta_{ 2 } = 77^{o}$. In this numerical analysis, the 
minimum value of the $\chi^{2}$, corresponding to the best fit, is $\chi^{2}=0.288$ and 
the resulting value of $\chi^{2}$ for degree of freedom is 
{\small $\frac{\chi^{2}_{min} }{ d.o.f. }=0.075$}. All numerical results of the fit are in 
very good agreement with the values of the moduli of the entries in the matrix 
$U_{ _{PMNS} }$  as given in Gonzalez-Garcia~\cite{GonzalezGarcia:2007ib}.

\section{The Mixing Angles}
In the standard PDG parametrization, the entries in the quark and lepton mixing matrices are 
parametrized in terms of the mixing angles and phases. Thus, the mixing angles are related 
to the observable moduli of quark (lepton) $ V_{ _{ CKM } } ( U_{ _{ PMNS } } )$ through the
relations:
\begin{equation}\label{angulosMezclas}
 \begin{array}{l}
  \sin^{2}{\theta_{12}^{q ( l ) } } = \frac{ \left| V_{us} \left(  U_{ e2 } \right) 
   \right|^{2} }{ 1 - \left| V_{ub} \left(  U_{ e3 } \right)  \right|^{2} }, \\\\ 
  \sin^{2} \theta_{23}^{  q ( l ) } =  \frac{ \left| V_{cb}  \left(  U_{ \mu 3 } \right) 
   \right|^{2} }{  1 - \left| V_{ub}  \left(  U_{ e3 } \right)\right|^{2} }, \\\\
  \sin^{2} \theta_{13}^{  q ( l ) } = \left| V_{ub}  \left(  U_{ e3 } \right) \right|^{2}.
 \end{array}  
\end{equation}
Then, theoretical expression for the quark mixing angles as functions of the quark mass 
ratios are readily obtained when the theoretical expressions for the modulii of
the entries in the $CKM$ mixing matrix, given in eqs.~(\ref{elem:ckm}) and~(\ref{Ds}), are 
substituted for $\left|V_{ij}\right|$ in the right hand side of eqs.~(\ref{angulosMezclas}). In 
this way,and keeping only the leading order terms, we get :
\begin{equation}
 \sin^{2}{\theta_{12}^{q ^{th} } } \approx  
 \frac{ \frac{ \widetilde{m}_{d} }{ \widetilde{m}_{s} } + \frac{ \widetilde{m}_{u} }{ 
 \widetilde{m}_{c} } - 2 \sqrt{ \frac{ \widetilde{m}_{u} }{ \widetilde{m}_{c} } 
 \frac{ \widetilde{m}_{d} }{ \widetilde{m}_{s} } } \cos{ \phi } }{  \left( 1 + 
 \frac{ \widetilde{m}_{u} }{ \widetilde{m}_{c} } \right) \left( 1 + \frac{ \widetilde{m}_{d} 
 }{ \widetilde{m}_{s} }  \right) },
\end{equation}
\begin{equation}
 \sin^{2} \theta_{23}^{  q ^{th} }   \approx   
 \frac{ \left( \sqrt{ \delta_{ u } } - \sqrt{ \delta_{ d  } } \right)^{2} }{ \left( 1 + 
 \frac{ \widetilde{m}_{u} }{ \widetilde{m}_{c} } \right) }, 
\end{equation}
\begin{equation}
 \sin^{2} \theta_{13}^{  q ^{th} }  \approx
 \frac{  \frac{ \widetilde{m}_{ u  } }{ \widetilde{m}_{ c } }   \left(  \sqrt{ \delta_{ u  } 
  } - \sqrt{ \delta_{ d  }  }  \right)^{2} }{  \left( 1 +  \frac{ \widetilde{m}_{u} 
  }{ \widetilde{m}_{c}} \right)  }.
\end{equation} 
Now, the numerical values of the quark mixing angles may be computed from 
eq.(\ref{elem:ckm}) and the numerical values of the parameters $\delta_{ u }$ and 
$\delta_{ d} $,eq. (\ref{X2:Qds}), and the CP violating phase $\phi=90^{o}$ obtained from 
$\chi^{2}$ fit of $\left| V_{ _{CKM} }^{ th } \right| $ to the experimentally determined 
values $\left| V_{ _{CKM} }^{ exp } \right| $. In this way we obtain 
\begin{equation}
 \theta_{12}^{q^{th}} = 13^{o},\quad \theta_{23}^{q^{th}} = 2.38^{o}, \quad  
 \theta_{13}^{q^{th}} = 0.19^{o}, 
\end{equation}
in very good agreement with the latest analysis of the experimental 
data~\cite{Mateu:2005wi}, see (\ref{PDGdatosang}). \\
The numerical values of the leptonic mixing angles are computed in a similar fashion. 
The theoretical expressions for the lepton mixing angles as funtion of the charged lepton  
and neutrino mass ratios are obtained from eqs (\ref{angulosMezclas}) when the theoretical
expressions for the modulii of the entries in the $PMNS$ mixing matrix, given in 
eqs.~(\ref{elem:pmns}) and (\ref{Ds}), are substituted for $\left| U_{ ij } \right| $ in the 
right hand side of eqs.(\ref{angulosMezclas}). If we keep only the leading orders terms, we 
obtain:
\begin{equation}\label{S12L}
 \begin{array}{l}
 \sin^{2}{\theta_{12}^{ l^{th} }} \approx 
 \frac{ 1 + \widetilde{m}_{\nu_{2}} - \delta_{ \nu } }{ \left( 1 + \widetilde{m}_{\nu_{2}} 
 \right) \left( 1 - \delta_{ \nu }  \right) \left( 1 + \frac{\widetilde{m}_{\nu_{1}} }{
 \widetilde{m}_{\nu_{2}} } \right)  \left( 1 + \frac{ \widetilde{m}_{e} }{ 
 \widetilde{m}_{\mu} } \right)}  \left \{ \frac{ \widetilde{m}_{\nu_{1}} }{ 
 \widetilde{m}_{\nu_{2}} } + 
 \right.  \\ \qquad \left.
  + \frac{ \widetilde{m}_{e} }{ \widetilde{m}_{\mu} } \left( 1 - \delta_{ \nu } \right)
 + 2 \sqrt{\frac{ \widetilde{m}_{\nu_{1}} }{ \widetilde{m}_{\nu_{2}} }
 \frac{ \widetilde{m}_{e} }{ \widetilde{m}_{\mu} } \left( 1 - \delta_{ \nu } \right) }
 \cos{ \Phi_{ _1 } }  \right \}, 
 \end{array}
\end{equation}
\begin{equation}\label{S23L}
 \sin^{ 2 } \theta_{23}^{ l^{th} } \approx 
 \frac{ \delta_{ \nu } + \delta_{ e } f_{ \nu2 } - \sqrt{ \delta_{ \nu } \delta_{ e } 
 f_{ \nu2 } } \cos \left( \Phi_{ _1 } - \Phi_{ _2 } \right) }{ \left( 1 +
 \frac{ \widetilde{m}_{e} }{ \widetilde{m}_{\mu} } \right) \left( 1 + 
 \widetilde{m}_{\nu_{2} } \right) },   
\end{equation}
\begin{equation}\label{S13L}
 \begin{array}{l}
 \sin^{ 2 } \theta_{13}^{ l^{th} } \approx  
 \frac{ \delta_{ \nu } }{ \left( 1 + \frac{ \widetilde{m}_{e} }{ \widetilde{m}_{\mu} }
 \right) \left( 1 + \widetilde{m}_{\nu_{2} } \right) } \left \{ \frac{ \widetilde{m}_{e} }{
 \widetilde{m}_{\mu} } + \frac{ \widetilde{m}_{\nu_{1} } \widetilde{m}_{\nu_{2}} }{ 
 \left( 1 - \delta_{ \nu } \right)} - 
 \right. \\ \left. 
 \qquad \qquad -2 \sqrt{ \frac{ \widetilde{m}_{e} }{
 \widetilde{m}_{\mu} } \frac{ \widetilde{m}_{\nu_{1}} \widetilde{m}_{ \nu_{2} } }{ 
 \left( 1 - \delta_{ \nu } \right) } } \cos \Phi_{ _1 } \right \}.   
 \end{array}
\end{equation}
From eqs.~(\ref{MsFsDs:leptones}) we have that 
$f_{ \nu2 } = 1 + \widetilde{m}_{\nu_{2}} - \delta_{ \nu }$.
The expressions quoted above are written in terms of the ratios of the lepton masses. When 
the well known values of the charged lepton masses, the values of the neutrino masses, 
eq.~(\ref{X2:Mnus}), the values of the delta parameters eq.~(\ref{X2:ds}) and  the values
of the Dirac CP violating phases obtained from $\chi^2$ fit in the lepton sector, are 
inserted in eqs.~(\ref{S12L})-(\ref{S13L}), we obtain the following  numerical values for
the mixing angles 
\begin{equation}
 \theta_{12}^{l^{th}} = 34.7^{o}, \quad  \theta_{23}^{l^{th}} = 43.6^{o}, \quad 
 \theta_{13}^{l^{th}} =  3.5^{o},
\end{equation}
which are in very good agreement with the latest experimental 
data~\cite{GonzalezGarcia:2007ib, GonzalezGarcia:2010er}.

\section{Quark-Lepton Complementarity}
The relations between mixing angles and the moduli of the entries of the mixing matrices given 
in eqs.~(\ref{angulosMezclas}) allow us to write the following identities:
\begin{equation}\label{QLC-M12}
  \tan{\left( \theta_{12}^{q} + \theta_{12}^{l } \right)} = 1 + \Delta_{12},
\end{equation}
 where
\begin{equation}\label{QLC-M12:Delta}
 \begin{array}{l}
 \Delta_{12} = \frac{  \left| V_{ us } \right| \left( \left| U_{ e1 } \right| +  
  \left| U_{ e2 } \right| \right) - \left| V_{ ud } \right| \left( \left| U_{ e1 } \right| -  
  \left| U_{ e2 } \right| \right) }{ \left| U_{ e1 } \right| 
  \left| V_{ ud } \right|- \left| U_{ e2 } \right| \left| V_{ us }\right|}.
 \end{array}
\end{equation} 
and
\begin{equation}\label{QLC-M23}
 \tan{ \left( \theta_{23}^{ q } + \theta_{23}^{ l }\right)} = 1 + \Delta_{23},
\end{equation}
where
\begin{equation}
 \begin{array}{l}
 \Delta_{23} = 
  \frac{  \left| V_{ cb } \right| \left( \left| U_{ \tau3 } \right| +  
  \left| U_{ \mu3 } \right| \right) - \left| V_{ tb } \right| \left( \left| U_{ \tau3 } 
  \right| -  \left| U_{ \mu3 } \right| \right) }{ \left| U_{ \tau3 } \right| 
  \left| V_{ tb } \right|- \left| U_{ \mu3 } \right| \left| V_{ cb }\right|}. 
  \end{array}
\end{equation} 
and
\begin{equation}\label{QLC-M13}
\begin{array}{l}
 \tan{ \left( \theta_{13}^{ q } + \theta_{13}^{ l }\right)} = 
 \frac{\left| V_{ ub } \right|\sqrt{ 1 - \left| U_{ e3 } \right|^{2}} + 
  \left| U_{ e3 } \right| \sqrt{ 1 -  \left| V_{ ub } \right|^{2}} }{ 
  \sqrt{ 1 -  \left| V_{ ub } \right|^{2}} \sqrt{ 1 - \left| U_{ e3 }
  \right|^{2}}  - \left| U_{ e3 } \right|  \left| V_{ ub } \right|}
\end{array}
\end{equation}
We notice that numerical values of $\Delta_{12}$ and $\Delta_{23}$ obtained from the 
experimentally determined $\left| V_{_{CKM}} \right|$ and $\left|U_{_{ PMNS } } \right|$ are 
much smaller than one, 
\begin{displaymath}
 \Delta_{12} \ll 1 \quad  \textrm{and} \quad \Delta_{23} \ll 1 ,
\end{displaymath} 
for this reason, the identities~(\ref{QLC-M12})-(\ref{QLC-M13}) are sometimes called
Quark Lepton Complementarity relations (QLC). 

The substitution  of expresions  (\ref{elem:ckm}) and (\ref{elem:pmns}) for the 
modulii of the elements of the mixing matrices $V_{_{CKM}}^{ th }$ and $U_{_{ PMNS } }^{ th }$,   
allows us express the  small terms $\Delta_{12}$ and $\Delta_{23}$ as funtions of the mass 
ratios of quarks and leptons. Then, the eqs.~(\ref{QLC-M12})-(\ref{QLC-M13}) take the following 
form: 
\begin{equation}\label{QLCTAN12}
  \tan{\left( \theta_{12}^{q^{th}} + \theta_{12}^{l^{th} } \right)} = 
  1 + \Delta_{12}^{^{th}} \left( 
   \frac{ \widetilde{m}_{u} }{ \widetilde{m}_{c} },
   \frac{ \widetilde{m}_{d} }{ \widetilde{m}_{s} },
   \frac{\widetilde{m}_{\nu_1} }{ \widetilde{m}_{\nu_2}},
   \frac{ \widetilde{m}_{e} }{ \widetilde{m}_{\mu} }
   \right),
\end{equation}
 where
\begin{widetext}
 \begin{equation}\label{DELTA}
  \begin{split}
  \begin{array}{l}
   \Delta_{12}^{  ^{th}  } \approx  
   \frac{
    \sqrt{ \frac{ \widetilde{m}_{d} }{ \widetilde{m}_{s} } + 
     \frac{ \widetilde{m}_{u} }{ \widetilde{m}_{c} } } 
      \left[ \sqrt{ \frac{ \widetilde{m}_{\nu_1} }{ \widetilde{m}_{\nu_2} } f_{ \nu2 } } 
      \left( 1 + \sqrt{ \frac{ \widetilde{m}_{e} }{ \widetilde{m}_{\mu} } 
      \frac{ \widetilde{m}_{\nu_2} }{ \widetilde{m}_{\nu_1} } \left( 1 - \delta_{ \nu } \right)
      } \right)  + \sqrt{ \left( 1 + \widetilde{m}_{\nu_2} \right) 
      \left( 1 - \delta_{ \nu } \right) } \right] 
      - \left[ \sqrt{ \left( 1 + \widetilde{m}_{\nu_2} \right) f_{ \nu1 } } - 
      \sqrt{ \frac{ \widetilde{m}_{\nu_1} }{ \widetilde{m}_{\nu_2} } f_{ \nu2 } } 
      \left( 1 + \sqrt{ \frac{ \widetilde{m}_{e} }{ \widetilde{m}_{\mu} } 
      \frac{ \widetilde{m}_{\nu_2} }{ \widetilde{m}_{\nu_1} } \left( 1 - \delta_{ \nu } \right)
      } \right) \right]  
     }{ \sqrt{ \left( 1 + \widetilde{m}_{\nu_2} \right) 
      \left( 1 - \delta_{ \nu } \right) } - 
      \sqrt{ \frac{ \widetilde{m}_{d} }{ \widetilde{m}_{s} } + 
     \frac{ \widetilde{m}_{u} }{ \widetilde{m}_{c} } } 
     \left( 1 + \sqrt{ \frac{ \widetilde{m}_{e} }{ \widetilde{m}_{\mu} } 
      \frac{ \widetilde{m}_{\nu_2} }{ \widetilde{m}_{\nu_1} } \left( 1 - \delta_{ \nu } \right)
      } \right)  }
  \end{array}
 \end{split}
\end{equation} 
\end{widetext} 
Here, rather than writing a lenghty but not very illuminating exact expresion, we give 
an approximate expression for $\Delta_{12}^{ ^{th} }$, whose numerical value differs from the 
exact expresion in $12\%$.
In the derivation of eq.~(\ref{DELTA}) from (\ref{QLC-M12:Delta}) we used the following 
approxinations
\begin{equation}
 \frac{ \left| V_{us}^{ ^{th} } \right| }{ \left| V_{ud}^{ ^{th} } \right| } \approx
 \sqrt{ \frac{ \widetilde{m}_{d} }{ \widetilde{m}_{s} } 
 + \frac{ \widetilde{m}_{u} }{ \widetilde{m}_{c} } } \approx 0.23152,
\end{equation}
which differs from the exact value in less than $1\%$, and
\begin{equation}
 \begin{array}{l}
  \frac{ \left| U_{ e2 }^{ ^{th} }\right| }{ \left| U_{ e1 }^{ ^{th} }\right|} \approx 
 \sqrt{ \frac{ \frac{\widetilde{m}_{\nu_1} }{ \widetilde{m}_{\nu_2} }  }{ 
 1 + \widetilde{m}_{\nu_2} } }
 \sqrt{ \frac{ 1 + \widetilde{m}_{\nu_2} - \delta_{ \nu } }{ 1 - \widetilde{m}_{\nu_1} - 
 \delta_{ \nu } } } \left\{ 1 +
 \right. \\ \left.
  \qquad \quad + \sqrt{ \frac{ \widetilde{m}_{e} }{ \widetilde{m}_{\mu}}  
 \frac{\widetilde{m}_{\nu_2} }{ \widetilde{m}_{\nu_1} } \left( 1 - \delta_{ \nu } \right) }
 \right\}
 \approx 0.688,
 \end{array}
\end{equation}
which differs from the exact value in less than $1\%$. \\
The identity~(\ref{QLCTAN12}) that defines 
 {\small $\Delta_{12}^{^{th}} \left(  \frac{ \widetilde{m}_{u} }{ \widetilde{m}_{c} },
   \frac{ \widetilde{m}_{d} }{ \widetilde{m}_{s} }, \frac{\widetilde{m}_{\nu_1} }{ 
   \widetilde{m}_{\nu_2}}, \frac{ \widetilde{m}_{e} }{ \widetilde{m}_{\mu} }
   \right)$} is frequently written in terms of the angle 
   $ \varepsilon^{ ^{th} }_{_{12}}$ that measures the desviation of 
   $\left( \theta_{12}^{q^{th}} + \theta_{12}^{l^{th} } \right)$ from 
   $\frac{\pi}{4}$. Then, eq.~(\ref{QLCTAN12}) may also be written as
\begin{equation}
  \tan{\left( \theta_{12}^{q^{th}} + \theta_{12}^{l^{th} } \right)} = 
  \tan{\left( \frac{\pi}{4} + \varepsilon^{ ^{th} }_{_{12}} \right)} = 1 + \Delta_{12}^{^{th}}.
\end{equation} 
From this expression, we get
\begin{equation}\label{corre:epsilon12}
 \varepsilon^{ ^{th} }_{_{12}} = \arctan \left\{ \frac{\Delta_{12}^{^{th}} }{
  2 + \Delta_{12}^{^{th}} } \right\}, \quad \left| \varepsilon^{ ^{th} }_{_{12}} \right|
  < \frac{\pi}{2} 
\end{equation}  
which given $ \varepsilon^{ ^{th} }_{_{12}}$ as funtion of the mass ratios of quarks 
and leptons. \\
Similarly,
\begin{equation}
  \tan{ \left( \theta_{23}^{ q ^{th} } +   \theta_{23}^{ l^{th} }\right)} = 1 +
  \Delta_{23}^{ ^{th} }\left( 
   \frac{ \widetilde{m}_{u} }{ \widetilde{m}_{c} },
   \frac{ \widetilde{m}_{d} }{ \widetilde{m}_{s} },
   \frac{\widetilde{m}_{\nu_1} }{ \widetilde{m}_{\nu_2}},
   \frac{ \widetilde{m}_{e} }{ \widetilde{m}_{\mu} }
   \right),
\end{equation}
where
\begin{widetext}
\begin{equation}
 \begin{split}
 \begin{array}{l}
   \Delta_{23}^{ ^{th} } \approx  
   \frac{ \left( \left[ \left( 1 + \frac{ \widetilde{m}_{e} }{ \widetilde{m}_{\mu} } \right) 
   \left( 1 + \widetilde{m}_{ \nu_2 } \right) - \delta_{ \nu } - \delta_{ e } f_{ \nu 2 }   
   \right]^{\frac{1}{2}} + \sqrt{ \delta_{ \nu } + \delta_{ e }   f_{ \nu 2 }  } \right) 
   \left( \sqrt{ 1 + \frac{ \widetilde{m}_{u} }{ \widetilde{m}_{c} } - \left( \sqrt{ 
   \delta_{ u} } - \sqrt{ \delta_{ d } } \right)^{2} } +   \left( \sqrt{ \delta_{ u} } - \sqrt{ 
   \delta_{ d } } \right) \right) }{\left[ \left( 1 + \frac{ \widetilde{m}_{e} }{ 
   \widetilde{m}_{\mu} } \right) 
   \left( 1 + \widetilde{m}_{ \nu_2 } \right) - \delta_{ \nu } - \delta_{ e } f_{ \nu 2 }   
   \right]^{\frac{1}{2}} \sqrt{ 1 + \frac{ \widetilde{m}_{u} }{ \widetilde{m}_{c} } - \left( 
   \sqrt{ \delta_{ u} } - \sqrt{ \delta_{ d } } \right)^{2} }-  \left(  \sqrt{ \delta_{ u} } - 
   \sqrt{ \delta_{ d } } \right)   \sqrt{ \delta_{ \nu } + 
   \delta_{ e }    f_{ \nu 2 }  }  } 
\end{array}
\end{split}
\end{equation}
Also,
\begin{equation}\label{QLC-T13}
 \begin{split}
  \begin{array}{l}
  \tan{ \left( \theta_{13}^{ q^{th} } + \theta_{13}^{ l^{th} }\right)}  \approx 
   \frac{ 
   \sqrt{ \frac{\widetilde{m}_{u} }{ \widetilde{m}_{c}} }
   \left( \sqrt{ \delta_{ u} } -  \sqrt{ \delta_{ d } } \right) 
   \left[ \left( 1 +  \frac{ \widetilde{m}_{e} }{ \widetilde{m}_{\mu} } \right) \left( 1 + 
   \widetilde{m}_{\nu_2} \right) - \delta_{\nu} \left( \sqrt{ \frac{ \widetilde{m}_{ \nu_1 } 
   \widetilde{m}_{ \nu_2 } }{ \left( 1 - \delta_{ \nu } \right) } } -  
   \sqrt{ \frac{ \widetilde{m}_{e} }{ \widetilde{m}_{\mu} } } \right)^{ 2 }  
   \right]^{ \frac{1}{2}} + 
   }{ 
   \sqrt{ 1 + \frac{ \widetilde{m}_{u} }{\widetilde{m}_{c}} - \frac{ \widetilde{m}_{u} 
    }{ \widetilde{m}_{c}}  \left( \sqrt{ \delta_{ u} } -\sqrt{ \delta_{ d } } \right)^{2} } 
    \left[  \left( 1 + \frac{ \widetilde{m}_{e} }{ \widetilde{m}_{\mu} } \right) \left( 1 + 
    \widetilde{m}_{ \nu_2 }  \right)  - \delta_{\nu} \left( \sqrt{ 
    \frac{ \widetilde{m}_{ \nu_1 } \widetilde{m}_{ \nu_2 } }{  \left( 1 - \delta_{ \nu } 
    \right) } } -  \sqrt{ \frac{ \widetilde{m}_{e} }{  \widetilde{m}_{\mu} } } \right)^{ 2 }  
    \right]^{\frac{1}{2}} - }  \\ 
  \frac{ 
   + \sqrt{ \delta_{ \nu } } \left( \sqrt{ \frac{ \widetilde{m}_{ \nu_1 } 
   \widetilde{m}_{ \nu_2 } }{ \left( 1 - \delta_{\nu} \right) } } -  \sqrt{ 
   \frac{ \widetilde{m}_{e} }{ \widetilde{m}_{\mu} } } \right) \sqrt{ 1 + 
   \frac{ \widetilde{m}_{u} }{ \widetilde{m}_{c} } - \frac{ \widetilde{m}_{u} }{  
   \widetilde{m}_{c} }\left( \sqrt{ \delta_{ u} } - \sqrt{ \delta_{ d } } \right)^{2} } }{ - 
   \sqrt{ \frac{\widetilde{m}_{u}}{\widetilde{m}_{c}} } \left( \sqrt{ \delta_{ u} } -
   \sqrt{ \delta_{ d } } \right)  \sqrt{ \delta_{ \nu } } \left(   \sqrt{ 
   \frac{ \widetilde{m}_{ \nu_1 } \widetilde{m}_{ \nu_2 } }{ 
   \left( 1 - \delta_{ \nu } \right) } } - \sqrt{ \frac{ \widetilde{m}_{e} }{ 
   \widetilde{m}_{\mu} } }  \right) }.
 \end{array}
 \end{split}
\end{equation}
\end{widetext}
After substitution of the numerical values of the mass ratios of quarks and leptons in 
eqs.~(\ref{DELTA})-(\ref{QLC-T13}), we obtain,
\begin{equation}
 \begin{array}{l}
   \Delta_{12}^{ ^{th} }= 0.1, \quad \Delta_{23}^{ ^{th} } = 3.23 \times 10^{-2}, \\\\
   \tan{ \left( \theta_{23}^{ q ^{th} } +   \theta_{23}^{ l^{th} }\right)} = 
   6.53 \times 10^{-2}.
 \end{array}
\end{equation}
Hence,
\begin{equation}
 \quad \theta_{12}^{q^{th}} + \theta_{12}^{l^{th}} = 45^{o} + 2.7^{o}. 
\end{equation}
\begin{equation}
   \theta_{23}^{ q^{th} } + \theta_{23}^{ l ^{th} }  = 45^{o}  + 1^{o},
\end{equation}
\begin{equation}
  \theta_{13}^{ q^{th} } + \theta_{13}^{ l^{th}} = 3.7^{o} .
\end{equation}
The equations (\ref{QLCTAN12}) and (\ref {DELTA}) are obtained from an exact analytical 
expression for $ \tan{\left( \theta_{12}^{q^{th}} + \theta_{12}^{l ^{th}} \right)} $  
as a funtion of the absolute values of the entries in the mixing matrices 
$V_{ _{CKM} }^{^{th}}$ and 
$U_{ _{PMNS} }^{^{th}}$, eqs (\ref{QLC-M12}) and (\ref{QLC-M12:Delta}).  
In eqs.~(\ref{elem:ckm}) and (\ref{elem:pmns}), the elements of the mixing matrices 
$V_{ _{CKM} }^{^{th}}$ and $U_{ _{PMNS} }^{^{th}}$ are given as exact, explicit analytical 
funtions of the quark and lepton mass ratios. Let us stress that these expressions are exact and 
valid for any possible values of the quark and lepton mass ratios. From~(\ref{DELTA}), it 
becomes evident that the small numerical value of $\Delta_{12}^{ ^{th} }$ is due to the partial 
cancellation of two large terms of almost the same magnitude but opposite sign appearing in the 
numerator of the expresion in the right hand side of the eq.~(\ref{DELTA}), namely,
{\small \begin{equation}
 \begin{array}{l}
  \sqrt{ \frac{ \widetilde{m}_{d} }{ \widetilde{m}_{s} } + 
     \frac{ \widetilde{m}_{u} }{ \widetilde{m}_{c} } } 
      \left[ \sqrt{ \frac{ \widetilde{m}_{\nu_1} }{ \widetilde{m}_{\nu_2} } f_{ \nu2 } } 
      \left( 1 + \sqrt{ \frac{ \widetilde{m}_{e} }{ \widetilde{m}_{\mu} } 
      \frac{ \widetilde{m}_{\nu_2} }{ \widetilde{m}_{\nu_1} } \left( 1 - \delta_{ \nu } \right)
      } \right)  +
      \right. \\  \left.
      \qquad + \sqrt{ \left( 1 + \widetilde{m}_{\nu_2} \right) 
      \left( 1 - \delta_{ \nu } \right) } \right] =0.287,
 \end{array}      
\end{equation} }
and
{\small \begin{equation}
 \begin{array}{l}
  \sqrt{ \left( 1 + \widetilde{m}_{\nu_2} \right) f_{ \nu1 } } - 
      \sqrt{ \frac{ \widetilde{m}_{\nu_1} }{ \widetilde{m}_{\nu_2} } f_{ \nu2 } } 
      \left( \; 1 + 
      \right. \\  \left.  \qquad
      + \sqrt{ \frac{ \widetilde{m}_{e} }{ \widetilde{m}_{\mu} } 
      \frac{ \widetilde{m}_{\nu_2} }{ \widetilde{m}_{\nu_1} } \left( 1 - \delta_{ \nu } \right)
      } \; \right) = 0.22.
 \end{array}       
\end{equation} }
The approximate numerical equality of these two expressions has its origin in the combined 
effect of the strong hierarchy of charged leptons and $u$ and  $d$-type quarks  which yields 
small and very small mass ratios, and the seesaw mechanism type~I which gives very small 
neutrino masses but relatively large neutrino mass ratios. \\

We may conclude that the so called Quark-Lepton Complementarity as expresed in~(\ref{QLCTAN12}) 
and~(\ref{DELTA}) is more than a numerical coincidence, it is the result of the combined 
effect of two factors:
\begin{enumerate}
 \item The strong mass hierarchy of the Dirac fermions which produces small and very small 
  mass ratios of $u$ and $d$-type quarks and charged leptons. The quark mass hierarchy is
  then reflected in a similar hierarchy of small and very small quark mixing angles.
 \item The normal seesaw mechanism type~I which gives very small masses to  the left-handed 
  Majorana neutrinos with relatively large values of the neutrino mass ratio   
  $m_{\nu_1}/m_{\nu_{2}}$ and allows for large $\theta_{12}^{l}$ and $\theta_{23}^{l}$ mixing 
  angles (see eqs.~(\ref{S12L})-(\ref{S13L})) .
\end{enumerate}
The two factors just mentioned contribute to numerator of $\Delta_{12}^{q^{th}}$ with two terms 
of almost equal magnitud but opposite sign. Hence, the small numerical value of 
$\Delta_{12}^{q^{th}}$ ocurring by partial cancellation of this two terms.

\section{The effective Majorana masses}
 The square of the magnitudes of the effective Majorana neutrino masses, 
 eq.(\ref{masa_eff.1}), are 
\begin{equation}\label{masa_eff.19}
\begin{array}{l}
  \left| \langle m_{ll} \rangle \right|^{2} =  \sum_{j=1}^{3} m_{ \nu_{j} }^{ 2 } 
  \left| U_{ lj } \right|^{ 4 }   +  2 \sum_{j<k}^{3} m_{ \nu_{j} } m_{ \nu_{k} } \times \\  
   \; \; 
  \times \left| U_{ lj } \right|^{ 2 }  \left| U_{ lk } \right|^{ 2 }  \cos 2\left( w_{lj} -
  w_{lk} \right), 
\end{array}
\end{equation}
where $ w_{lj} = \arg \left \{   U_{ lj } \right \}$; this term includes phases of 
both types, Dirac and Majorana.

The theoretical expression for the squared magnitud of the effective Majorana neutrino mass 
of electron neutrino, written in terms of the ratios of the lepton masses, is:
\begin{equation}\label{eff-ele}
 \begin{array}{l}
  \left| \langle m_{ee} \rangle \right|^{2} \approx  
  \frac{ 1 }{ \left( 1 + \frac{ \widetilde{m}_{ e } }{ \widetilde{m}_{ \mu }} \right)^{ 2 }
  \left( 1 +  \frac{\widetilde{m}_{\nu_1}}{\widetilde{m}_{\nu_2}} \right)^{ 2 }  } \left 
 \{  m_{\nu_1}^{ 2 }  \left( 1 - \right. \right. \\ \left. \left. 
 -4 \sqrt{ \frac{ \widetilde{m}_{ e } }{  \widetilde{m}_{ \mu }}  \frac{ 
 \widetilde{m}_{\nu_1} }{ \widetilde{m}_{\nu_2} }  \left( 1 - \delta_{ \nu } \right) }  
 \right) + \frac{ m_{\nu_2}^{ 2 } f_{\nu2}^2  }{  \left( 1 +  \widetilde{m}_{\nu_2} 
 \right)^{2}  \left( 1 - \delta_{ \nu } \right)^{2} }  
 \frac{ \widetilde{m}_{\nu_1} }{  \widetilde{m}_{\nu_2} }  \left( \frac{ 
 \widetilde{m}_{\nu_1} }{ \widetilde{m}_{\nu_2} }
   \right. \right. \\  \left. \left.      
 +  4 \sqrt{ \frac{ \widetilde{m}_{ e } }{ \widetilde{m}_{ \mu } }   \frac{ 
 \widetilde{m}_{\nu_1} }{ \widetilde{m}_{\nu_2} }  
 \left( 1 - \delta_{ \nu } \right) }  + 6 \frac{ \widetilde{m}_{ e } }{ 
 \widetilde{m}_{ \mu }}  \left( 1 -  \delta_{ \nu } \right)  \right)
    \right. \\  \left. 
 + 2 \frac{ m_{\nu_1} m_{\nu_3} \delta_{ \nu} }{ \left( 1 +  \widetilde{m}_{\nu_2} \right) }  
 \left( 1 +  \frac{ \widetilde{m}_{\nu_1} }{ \widetilde{m}_{\nu_2} }  \right)  
 \left( \sqrt{ \frac{ \widetilde{m}_{\nu_1} \widetilde{m}_{\nu_2}}{  \left( 1 
 - \delta_{ \nu } \right) } }  - \sqrt{ \frac{ \widetilde{m}_{ e } }{ 
 \widetilde{m}_{ \mu }} }  \right)^{ 2 } \times \right. \\ \left. \times 
 \cos 2( w_{e1} - w_{e3} )   +2 \frac{ m_{\nu_1} m_{\nu_2} f_{\nu2} }{ \left( 1 +  
 \widetilde{m}_{\nu_2} \right)  \left( 1 - \delta_{ \nu } \right) } \left( 
 \frac{\widetilde{m}_{\nu_1}}{\widetilde{m}_{\nu_2}} 
  \right. \right. \\  \left.\left.
 +   2 \left( 1 -  \frac{\widetilde{m}_{\nu_1}}{\widetilde{m}_{\nu_2}} \right)  
 \sqrt{ \frac{\widetilde{m}_{\nu_1}}{\widetilde{m}_{\nu_2}}  
 \frac{ \widetilde{m}_{ e } }{ \widetilde{m}_{ \mu }}  \left( 1 - \delta_{ \nu } \right) } 
 \right) \cos 2( w_{e1} - w_{e2} ) 
  \right. \\  \left.
 + 2 \frac{ m_{\nu_2} m_{\nu_3} f_{\nu2} \delta_{ \nu } }{  \left( 1 +  
 \widetilde{m}_{\nu_2} \right)^{2}   \left( 1 - \delta_{ \nu } \right)^{2} }  
 \left( 1 +  \frac{ \widetilde{m}_{\nu_1} }{ \widetilde{m}_{\nu_2} } \right)  
 \left( 2 \widetilde{m}_{\nu_1}  \widetilde{m}_{\nu_2} 
 \right. \right. \\  \left.\left.
 + \sqrt{ \frac{ \widetilde{m}_{ e } }{  \widetilde{m}_{ \mu }} 
 \frac{\widetilde{m}_{\nu_1}}{\widetilde{m}_{\nu_2}}   \left( 1 - \delta_{ \nu } \right)  } 
 \right) \cos 2( w_{e2} - w_{e3} )  
\right\}
\end{array}
\end{equation}
where $w_{ e2 } \approx \beta_{1} $ and 
\begin{equation}
 w_{ e1 } = \arctan \left\{  - \frac{ \sqrt{ \frac{\widetilde{m}_{\nu_1}}
 {\widetilde{m}_{\nu_2}}  \frac{\widetilde{m}_{e}
 }{\widetilde{m}_{\mu}} \delta_{e} \delta_{ \nu }   f_{\nu2} } }{ \sqrt{ \left( 1 
 - \delta_{ \nu } \right) } +  \sqrt{ \frac{\widetilde{m}_{\nu_1} }{ 
 \widetilde{m}_{\nu_2} } \frac{\widetilde{m}_{e}}{\widetilde{m}_{\mu}} } \left( 1 
 - \delta_{ \nu } \right) } \right \}, 
 \end{equation}
\begin{equation}
 \begin{array}{l} 
   w_{ e3 } \approx \arctan 
  \left\{ \frac{ \sqrt{ \frac{ \widetilde{m}_{e} }{ \widetilde{m}_{\mu} } \delta_{ e } 
  f_{ \nu 2} \left( 1 - \delta_{ \nu } \right) } + }{ - \sqrt{ \frac{ \widetilde{m}_{e} 
  }{ \widetilde{m}_{\mu} } \delta_{ e } f_{ \nu 2} \left( 1 - \delta_{ \nu } \right) } 
  \tan \beta_{ 2 } + }     
  \right. \\ \left. \qquad \quad 
 \frac{+\sqrt{ \delta_{ \nu } }  \left( \sqrt{ \widetilde{m}_{\nu_1} \widetilde{m}_{\nu_2} } 
 - \sqrt{ \frac{ \widetilde{m}_{e} }{ \widetilde{m}_{\mu} } \left( 1 - \delta_{ \nu } 
 \right) } \right) \tan \beta_{2} }{ + \sqrt{ \delta_{ \nu } }  \left( \sqrt{
  \widetilde{m}_{\nu_1} \widetilde{m}_{\nu_2} } -  \sqrt{ \frac{\widetilde{m}_{e} }{
  \widetilde{m}_{\mu} }   \left( 1 - \delta_{ \nu } \right) } \right)   } \right \}.
 \end{array}
\end{equation}
In a similar way, the theoretical expression for the squared magnitud of the effective 
Majorana neutrino mass of the  muon neutrino is:
\begin{equation}\label{eff-mu}
 \begin{array}{l}
  \left| \langle m_{\mu \mu} \rangle \right|^{2} \approx  \frac{1}{ \left( 1 + 
  \frac{ \widetilde{m}_{e} }{ \widetilde{m}_{\mu} } \right)^{2}  \left( 1 + 
  \frac{ \widetilde{m}_{\nu_1} }{ \widetilde{m}_{\nu_2} }  \right)^{2}  \left( 1
  + \widetilde{m}_{\nu_2} \right) }  \left \{  \frac{ m_{\nu_3}^{ 2 } }{ \left( 1 + 
  \widetilde{m}_{\nu_2} \right) }   \right. \\ \left.  
  \left( 1 + \frac{\widetilde{m}_{\nu_1}}{\widetilde{m}_{\nu_2}}  \right)^{2}  
  \left( \delta_{ \nu } + 2  \delta_{ e }  f_{\nu2} \right) + \frac{ m_{\nu_2}^{ 2 } }{ 
  \left( 1 + \widetilde{m}_{\nu_2} \right)  \left( 1 - \delta_{ \nu } \right) }  
  \left( 1 -  \delta_{ \nu } 
  \right. \right. \\  \left. \left.
  -4 \sqrt{ \frac{\widetilde{m}_{e}}{\widetilde{m}_{\mu}}  \frac{ \widetilde{m}_{\nu_1} 
  }{ \widetilde{m}_{\nu_2}}  \left( 1 - \delta_{ \nu } \right) }  + 6 
  \frac{ \widetilde{m}_{e} }{ \widetilde{m}_{\mu}}  \frac{\widetilde{m}_{\nu_1} }{ 
  \widetilde{m}_{\nu_2}} \right) +2 m_{\nu_1}m_{\nu_2} f_{\nu2}  
   \right. \\  \left.
  \left(  \frac{\widetilde{m}_{\nu_1}}{\widetilde{m}_{\nu_2}}  \left( 1 - \delta_{ \nu }
  \right)  +2 \sqrt{  \frac{\widetilde{m}_{\nu_1}}{\widetilde{m}_{\nu_2}} 
  \frac{\widetilde{m}_{e}}{\widetilde{m}_{\mu}} 
 \left( 1 - \delta_{ \nu } \right) }  \left( 1 - \frac{ \widetilde{m}_{\nu_1} }{ 
 \widetilde{m}_{\nu_2}}  \right)   \right) 
   \right. \\  \left.\cos 2(w_{\mu 1} -w_{ \mu 2} )
  + 2 m_{\nu_1}m_{\nu_3}  \left( 1 + \frac{\widetilde{m}_{\nu_1}}{\widetilde{m}_{\nu_2}}  
  \right)  \left( 2 \delta_{ \nu } 
 \times   \right. \right. \\  \left.\left.  \times \sqrt{ \frac{\widetilde{m}_{\nu_1} }{ 
 \widetilde{m}_{\nu_2}}  \frac{\widetilde{m}_{e}}{\widetilde{m}_{\mu}} \left( 1 - 
 \delta_{ \nu } \right) }  +  \frac{\widetilde{m}_{\nu_1}}
  {\widetilde{m}_{\nu_2}}  \left( 1 - \delta_{ \nu } \right)  \left( \delta_{ \nu } +   
  \delta_{ e }  f_{\nu2} \right)\right)
 \right. \\  \left.  
 \cos 2(w_{\mu 1} -w_{ \mu 3} ) +2 \frac{ m_{\nu_2}m_{\nu_3} f_{\nu2}  }{ \left( 1 + 
 \widetilde{m}_{\nu_2} \right)   \left( 1 - \delta_{ \nu } \right) }  \left( 1 + 
 \frac{\widetilde{m}_{\nu_1}}{\widetilde{m}_{\nu_2}}  \right) 
  \right.  \\  \left. 
  \left(  \left( 1 - \delta_{ \nu } \right) \left( \delta_{ \nu } +  \delta_{ e }  f_{\nu2} 
  \right) - 2 \delta_{ \nu }  \sqrt{ \frac{\widetilde{m}_{\nu_1}}{\widetilde{m}_{\nu_2}}  
  \frac{ \widetilde{m}_{e} }{ \widetilde{m}_{\mu} }  \left( 1 - \delta_{ \nu } \right) }   
  \right) 
 \right.  \\  \left. \cos 2(w_{\mu 2} -w_{ \mu 3} )  \right \}
 \end{array} 
\end{equation}
 where 
 \begin{equation}
 w_{ \mu 1 } \approx \arctan \left \{ 
 \frac{ \sqrt{ \frac{\widetilde{m}_{\nu_1}}{\widetilde{m}_{\nu_2} } \delta_{ e } 
 \delta_{ \nu }  f_{\nu2} } }{  \sqrt{ \frac{\widetilde{m}_{e}}{\widetilde{m}_{\mu}}   
 \left( 1 - \delta_{ \nu } \right) } +
 \sqrt{ \frac{\widetilde{m}_{\nu_1}}{\widetilde{m}_{\nu_2} }  }  \left( 1 - \delta_{ \nu } 
 \right) } \right \} ,
 \end{equation}
 and
 \begin{equation}
 w_{ \mu 2} \approx \arctan \left \{ \frac{ \sqrt{ f_{\nu2} } \tan \beta_{1} + \sqrt{ 
 \delta_{e} \delta_{ \nu } }  }{ \sqrt{ f_{\nu2} }  - \sqrt{ \delta_{e} \delta_{ \nu } } 
 \tan \beta_{1} } \right \}, 
\end{equation}
\begin{equation} 
   w_{ \mu 3 } \approx
   \arctan \left \{  \frac{  \tan \beta_{2} - \sqrt{ f_{ \nu2 } }  }{ 1 + 
 \sqrt{ f_{ \nu2 } } \tan \beta_{2} } \right \}.
\end{equation}
From these expresions and the numerical values of the neutrinos masses given in 
eq.~(\ref{X2:Mnus}), we obtain the following expresions for effective Majorana masses with 
the phases as a free parameters:
{\small \begin{equation}
\begin{array}{l} 
 \left| \langle m_{ee} \rangle \right|^{2} \approx 
 \left \{ 9.41 + 8.29 \cos ( 1^{o} - 2\beta_{1} ) + 4.3  \cos ( 1^{o} - 2w_{e3} ) 
  \right. \\ \left. 
  + 4.31 \cos 2( \beta_{1} - w_{e3} )  
\right \} \times 10^{-6}~\textrm{eV}^2
\end{array}
\end{equation} }
where
\begin{equation}
 w_{e3} = \arctan \left \{ \frac{0.15 \tan \beta_{2} - 0.013 }{ 
 0.15 + 0.013 \tan \beta_{2} } \right \}. 
\end{equation}
Similarly,
\begin{equation}
\begin{array}{l} 
 \left| \langle m_{\mu \mu} \rangle \right|^{2} \approx 
 \left \{ 4.8 + 0.17 \cos 2( 44^{o} - w_{\mu 2}   ) 
  \right. \\ \left. 
  + 1.8  \cos 2( w_{\mu 2} - w_{\mu 3} ) 
 \right \} \times 10^{-4}~\textrm{eV}^2
\end{array}
\end{equation}
where
\begin{equation}
 w_{ \mu 2} \approx \arctan \left \{ 
 \frac{ 0.65 \tan \beta_{1} + 0.13 }{ 0.65 - 0.13 \tan \beta_{1} } \right \}, 
\end{equation}
\begin{equation} 
   w_{ \mu 3 } \approx
   \arctan \left \{  \frac{  \tan \beta_{2} - 0.13  }{ 1 + 
 0.13 \tan \beta_{2} } \right \}.
\end{equation}
In order to make a numerical estimate of the effective Majorana neutrinos masses 
$\left| \langle m_{ee} \rangle \right|$ and $\left| \langle m_{\mu \mu} \rangle \right|$,
we used the following values for the Majorana phases $\beta_{1 }=-1.4^{o}$ and 
$\beta_{ 2 }=77^{o}$ obtained by maximizing the rephasing invariants $S_{1}$ and $S_{2}$,
eq.~(\ref{valor:S1S2}). Then, the numerical value of the effective Majorana neutrino 
masses are:
 \begin{equation}
  \left| \langle m_{ee} \rangle \right| \approx 4.6  \times 10^{ -3 }~\textrm{eV} ,  \quad 
  \left| \langle m_{\mu \mu} \rangle \right| \approx 2.1  \times 10^{ -2 }~\textrm{eV} .
 \end{equation}
 These numerical values are consistent with the very small experimentally determined upper 
 bounds for the reactor neutrino mixing angle  
 $\theta_{13}^{l}$~\cite{PhysRevLett.101.141801}.

 \section{Conclusions}
 In this communication, we outlined a unified treatment of masses and mixings of quarks and 
 leptons in which the left-handed Majorana neutrinos acquire their  masses via the type I
 seesaw mechanism, and the mass matrices of all Dirac fermions have a similar form with 
 four texture zeroes and a normal hierarchy. Then, the mass matrix of the left-handed Majorana 
 neutrinos  also has a texture with four zeros. In this scheme, we derived exact, explicit
 expressions for the  Cabibbo ($\theta_{12}^{q}$) and solar $(\theta_{12}^{l})$ mixing 
 angles as functions of the quark and lepton masses, respectively. The so called Quark-Lepton 
 Complementarity relation takes the form, 
 \begin{equation}
  \theta_{12}^{q^{th}} + \theta_{12}^{l^{th}} = 45^{o} + \varepsilon^{ ^{th} }_{_{12}}.
 \end{equation}
 The correction term, $\varepsilon^{ ^{th} }_{_{12}}$, is an explicit function of the ratios of  
 quark and  lepton masses, given in eq.~(\ref{corre:epsilon12}), which reproduces the 
 experimentally determined  value,
 \begin{equation}
 \varepsilon_{12}^{^{exp}} \approx 2.7^{o}, 
 \end{equation}
 when the numerical values of the quark and lepton masses are substituted 
 in~(\ref{corre:epsilon12}).

 Three essential ingredients are needed to explain the correlations implicit in the small 
 numerical value of $\varepsilon^{ ^{th} }_{_{12}}$:
 \begin{enumerate}
  \item The strong hierarchy in the mass spectra of the quarks and charged leptons, realized  
  in our scheme through the explicit breaking of the $S_{3}$ flavor symmetry in the  
  mass matrices with four texture zeroes, explains the resulting small or very small  
  quark mixing angles, the very small charged lepton mass ratios explain the  very small 
  value of $\theta_{13}^{l}$.
  \item The normal seesaw mechanism that gives very small masses to the left-handed Majorana 
  neutrinos with relatively large values of the neutrino mass ratio $m_{\nu_1}/m_{\nu_{2}}$ 
  and allows for large $\theta_{12}^{l}$ and $\theta_{23}^{l}$ mixing angles.
  \item The assumption of a normal hierarchy for the masses of the Majorana neutrinos.
 \end{enumerate} 
 
 \acknowledgments{ We thank Dr. Myriam Mondrag\'on for many inspiring discussions on this 
 exciting problem. This work was partially supported by CONACyT Mexico under Contract No. 
 51554-F and 82291, and DGAPA-UNAM Contract No. PAPIIT IN112709.}
%%%%%%%%%%%%%%%%%%%%%%%%%%%%%%%%%%%%%%%%%%%
%% The following lines show an example how to produce a bibliography
%% without the help of the BibTeX program. This could be used instead
%% of the above.
%%%%%%%%%%%%%%%%%%%%%%%%%%%%%%%%%%%%%%%%%%%

%\bibliographystyle{}
%\bibliography{Bibliografia}

\end{document}